\documentclass[reprint,superscriptaddress,preprintnumbers,frontmatterverbose,amsmath,amssymb,aps,prd,floatfix,nofootinbib]{revtex4-2}

\usepackage{graphicx}
\usepackage{dcolumn}
\usepackage{bm}
\usepackage{hyperref}
\usepackage{orcidlink}
\usepackage{xcolor,amsmath}
\usepackage{flushend}
\usepackage{multirow}
\usepackage{subcaption}
\usepackage{float}
\usepackage{cleveref}
\usepackage{stfloats}

\usepackage{rotating}

\def \ra{\rightarrow}
\def \Kb{{\bar K^0}}

\def \beq{\begin{equation}}
\def \eeq{\end{equation}}
\def \bea{\begin{eqnarray}}
\def \eea{\end{eqnarray}}
\def \bma{\begin{matrix}}
\def \ema{\end{matrix}}

\def \({\left(}
\def \){\right)}
\def \[{\left[}
\def \]{\right]}
\def \nn{\nonumber}

\def \Z2{\mathbb{Z}_2}

\def\tT{{\widetilde{T}}}
\def\tC{{\widetilde{C}}}
\def\tA{{\widetilde{A}}}

\def\tPuc{{\widetilde{P_{uc}}}}

\def\tPAuc{{\widetilde{PA_{uc}}}}
\def\btopik{B\to\pi K}

\newcommand\varpm{\mathbin{\vcenter{\hbox{%
  \oalign{\hfil$\scriptstyle+$\hfil\cr
          \noalign{\kern-.3ex}
          $\scriptscriptstyle({-})$\cr}%
}}}}

\allowdisplaybreaks

\begin{document}

\title{\boldmath Anomalies in Hadronic $B$ Decays: an Update}

\author{Bhubanjyoti Bhattacharya\,\orcidlink{0000-0003-2238-321X}}
\email{bbhattach@ltu.edu}
\affiliation{Department of Natural Sciences, Lawrence Technological University, Southfield, MI 48075, USA}

\author{Marianne Bouchard\,\orcidlink{0009-0005-2885-7473}}
\email{marianne.bouchard.5@umontreal.ca}
\affiliation{Physique des Particules, Universit\'e de Montr\'eal, 1375 Avenue Th\'er\`ese-Lavoie-Roux, Montr\'eal, QC, Canada  H2V 0B3}

\author{Luke Hudy\,\orcidlink{0009-0009-5054-793X}}
\email{lhudy@ltu.edu}
\affiliation{Department of Natural Sciences, Lawrence Technological University, Southfield, MI 48075, USA}

\author{Alexandre Jean\,\orcidlink{0009-0005-2354-3299}}
\email{alexandre.jean.1@umontreal.ca}
\affiliation{Physique des Particules, Universit\'e de Montr\'eal, 1375 Avenue Th\'er\`ese-Lavoie-Roux, Montr\'eal, QC, Canada  H2V 0B3}

\author{David London\,\orcidlink{0000-0002-4407-5624}}
\email{london@lps.umontreal.ca}
\affiliation{Physique des Particules, Universit\'e de Montr\'eal, 1375 Avenue Th\'er\`ese-Lavoie-Roux, Montr\'eal, QC, Canada  H2V 0B3}

\author{Christopher MacKenzie\,\orcidlink{0009-0005-2275-9573}} 
\email{cmackenzi@ltu.edu}
\affiliation{Department of Natural Sciences, Lawrence Technological University, Southfield, MI 48075, USA}

\date{\today}

\preprint{UdeM-GPP-TH-25-306}

\begin{abstract}
Recently, $B\to PP$ decays ($B = \{B^0, B^+, B_s^0\}$, $P = \{ \pi, K \}$) were analyzed under the assumption of flavor SU(3) symmetry (SU(3)$_F$). Although the individual fits to $\Delta S=0$ or $\Delta S=1$ decays are good, it was found that the combined fit is very poor: there is a $3.6\sigma$ disagreement with the SU(3)$_F$ limit of the standard model (SM$_{\rm{SU(3)}_F}$). One can remove this discrepancy by adding SU(3)$_F$-breaking effects, but 1000\% SU(3)$_F$ breaking is required. In this paper, we extend this analysis to include decays in which there is an $\eta$ and/or $\eta'$ meson in the final state. We now find that the combined fit exhibits a $4.1\sigma$ discrepancy with the SM$_{\rm{SU(3)}_F}$, and 1000\% SU(3)$_F$-breaking effects are still required to explain the data. These results are rigorous, group-theoretically -- no theoretical assumptions have been made. But when one adds some theoretical input motivated by QCD factorization, the discrepancy with the SM$_{\rm{SU(3)}_F}$ grows to $4.9\sigma$.
\end{abstract}

\maketitle

\section{Introduction}

There have been a number of measurements of observables in hadronic $B$ decays that disagree with the predictions of the standard model (SM). One of these -- the $\btopik$ puzzle (see Refs.~\cite{Beaudry:2017gtw, Bhattacharya:2021shk} and references therein) -- has been around for more than 20 years. Other, more recent examples can be found in Refs.~\cite{Alguero:2020xca, Bhattacharya:2022akr, Amhis:2022hpm}. 

In Ref.~\cite{Berthiaume:2023kmp}, for which three of the present authors (BB, AJ, DL) are co-authors, $B\to PP$ decays are examined under the assumption of flavor SU(3) symmetry [SU(3)$_F$]. Here $B = \{B^0, B^+, B_s^0\}$ and the pseudoscalar $P = \{ \pi, K \}$.  Fits to the latest data are performed, and the results are extremely intriguing. The fits to $\Delta S=0$ or $\Delta S=1$ decays individually are good. However, the combined fit is very poor: there is a $3.6\sigma$ disagreement with the SU(3)$_F$ limit of the standard model (SM$_{\rm{SU(3)}_F}$). 

This discrepancy can be removed by adding SU(3)$_F$-breaking effects, but 1000\% SU(3)$_F$ breaking is required (i.e., parameters that are equal in the SU(3)$_F$ limit must now differ by a factor of $\sim$10). These results are rigorous, group-theoretically -- no dynamical theoretical assumptions have been made. But when one adds an assumption motivated by QCD factorization, the discrepancy with the SM$_{\rm{SU(3)}_F}$ grows to $4.4\sigma$. It is perhaps too soon to claim that new physics must be present, but it is clear that something unexpected is going on here.

The reason that we restrict the pseudoscalar meson $P$ to be a pion or a kaon is that these seven particles (three $\pi$s, four $K$s) are members of the SU(3)$_F$ meson octet ${\bf 8}$, and can be considered to be identical particles in the SU(3)$_F$ limit. But there is an eighth member of the octet that has not been included, the $\eta_8$. This is not a physical particle -- it mixes with the SU(3)$_F$ singlet $\eta_1 = {\bf 1}$ to form the $\eta$ and $\eta'$ pseudoscalar mesons. In this paper, we extend the above analysis to include these missing particles. That is, we consider $B\to PP$ decays, but here $P$ includes all the pseudoscalar mesons, $\{ \pi, K, \eta, \eta' \}$. Previously, the final state was purely $\boldsymbol{8\otimes8}$; now we also include $\boldsymbol{8\otimes1}$ and $\boldsymbol{1\otimes1}$ final states. 

The purpose of this study is to examine what happens to the discrepancy with the SM$_{\rm{SU(3)}_F}$ when these additional decays are included. Also, although all members of the ${\bf 8}$ can be considered to be identical particles under SU(3)$_F$, the singlet $\eta_1$ is different from these particles. So what happens to the size of SU(3)$_F$ breaking in this case? As we will see, even when decays to final states involving $\eta$/$\eta'$ are included, the global fit remains very poor --  the discrepancy with the SM$_{\rm{SU(3)}_F}$ reaches $4.1\sigma$, and 1000\% SU(3)$_F$ breaking is still indicated.

The main mathematical tool used in this study is basically the Wigner-Eckart theorem within SU(3)$_F$. That is, all $B\to PP$ amplitudes are expressed as functions of SU(3)$_F$ reduced matrix elements (RMEs). In the fits, the RMEs are taken to be the unknown theoretical parameters. By fitting to the data, we determine what values of the RMEs best reproduce the data, and we also compute how good (or bad) this fit is.

One difficulty with using RMEs is that it is necessary to compute the SU(3)$_F$ Clebsch-Gordan coefficients, which can be tricky. For this reason, an alternative method is usually used, that of topological quark diagrams \cite{Gronau:1994rj, Gronau:1995hn}. It has been shown that, for $B\to PP$ decays with $P = \{ \pi, K \}$, the set of these diagrams is equivalent to RMEs. But the advantage of diagrams is that, when one writes the $B\to PP$ amplitudes in terms of diagrams, knowledge of the SU(3)$_F$ Clebsch-Gordan coefficients is not required. Also, from a theoretical point of view, one can estimate the relative sizes of different diagrams, whereas one cannot do this with RMEs. In the analysis of Ref.~\cite{Berthiaume:2023kmp}, diagrams were used.

In this paper, we present the analysis in terms of both RMEs and diagrams. There are several reasons for this. First, while the equivalence of diagrams and RMEs has been shown for the $\boldsymbol{8\otimes8}$ final states, it has not been demonstrated for the $\boldsymbol{8\otimes1}$ and $\boldsymbol{1\otimes1}$ final states. Here we present the relations between diagrams and RMEs for these final states. Also, there are a number of subtle relationships among the RMEs -- the number of independent RMEs is actually smaller than what is found by naive counting (which is often used in the literature). A similar conclusion holds for the diagrams, and one must properly take into account these relations in performing the fits. In order to identify the independent RMEs and diagrams, it is useful to perform the derivation in the two different (but equivalent) ways.

We begin in Sec.~2 with a discussion of the $B\to PP$ amplitudes with $P = \{ \pi, K, \eta, \eta' \}$, for the $\boldsymbol{8\otimes8}$, $\boldsymbol{8\otimes1}$ and $\boldsymbol{1\otimes1}$ final states.  This is done from both points of view, RMEs and diagrams. We also show the equivalence of RMEs and diagrams for all three classes of final states. We perform the fits in Sec.~3, first updating the $\boldsymbol{8\otimes8}$ fits of Ref.~\cite{Berthiaume:2023kmp}, and then adding the $\boldsymbol{8\otimes1}$ and $\boldsymbol{1\otimes1}$ decays. In Sec.~4, we examine the effect of adding certain well-motivated theoretical assumptions to the fits. We conclude in Sec.~5.

\section{\boldmath $B\to PP$ amplitudes}
\label{Sec:BPPamps}

There are nine pseudoscalar ($P$) mesons. The SU(3)$_F$ octet ${\bf 8}$ contains $\{\pi^\pm, \pi^0, K^\pm, K^0, {\bar K}^0, \eta_8 \}$, while the singlet ${\bf 1}$ is $\eta_1$. (The physical $\eta$ and $\eta'$ mesons are linear combinations of $\eta_8$ and $\eta_1$.) There are a total of 17 $\Delta S = 0$ and 17 $\Delta S = 1$ $B \to PP$ decays, separated into three categories for the $PP$ final state: $\boldsymbol{8\otimes8}$ (24 decays), $\boldsymbol{8\otimes1}$ (8 decays) and $\boldsymbol{1\otimes1}$ (2 decays). These are listed in Table \ref{tab:BPPdecays}.

\begin{table}
\begin{center}
\begin{tabular}{l c|c l}
    \multicolumn{4}{c}{$\boldsymbol{8\otimes 8}$}\\
    \hline
    \multicolumn{2}{c|}{$\Delta S = 0$} &
    \multicolumn{2}{c}{$\Delta S = 1$}\\
    \hline\hline
    \vspace{-0.3cm}&&&\\
    $B^+\rightarrow {\bar K^0}K^+$ &&& $B^+\rightarrow \pi^+K^0$\\
    $B^+\rightarrow \pi^+\pi^0$ &&& $B^+\rightarrow \pi^0K^+$\\
    $B^+\rightarrow \eta_8 \pi^+$ &&& $B^+\rightarrow \eta_8 K^+$\\
    \hline
    \vspace{-0.3cm}&&&\\
    $B^0\rightarrow K^0{\bar K^0}$ &&& $B^0_s\rightarrow K^0{\bar K^0}$ \\
    $B^0\rightarrow\pi^+\pi^-$ &&&$B^0_s\rightarrow\pi^+\pi^-$ \\
    $B^0\rightarrow \pi^0\pi^0$&&& $B^0_s\rightarrow\pi^0\pi^0$\\
    $B^0\rightarrow K^+K^-$ &&& $B^0_s\rightarrow K^+K^-$\\
    $B^0\rightarrow \pi^0\eta_8$ &&& $B^0_s\rightarrow\pi^0\eta_8$\\
    $B^0\rightarrow \eta_8\eta_8$&&& $B^0_s\rightarrow \eta_8\eta_8$\\
    \hline
    \vspace{-0.3cm}&&&\\
    $B^0_s\rightarrow\pi^+K^-$&&&$B^0\rightarrow\pi^-K^+$\\
    $B^0_s\rightarrow\pi^0{\bar K^0}$&&&$B^0\rightarrow \pi^0 K^0$\\
    $B^0_s\rightarrow\eta_8{\bar K^0}$&&&$B^0\rightarrow \eta_8 K^0$\\
    \\
    \multicolumn{4}{c}{$\boldsymbol{8\otimes 1}$}\\
    \hline
    \multicolumn{2}{c|}{$\Delta S = 0$} &
    \multicolumn{2}{c}{$\Delta S = 1$}\\
    \hline\hline
    \vspace{-0.3cm}&&&\\
    $B^+\rightarrow \eta_1\pi^+$ &&& $B^+\rightarrow \eta_1 K^+$\\
    \hline
    \vspace{-0.3cm}&&&\\
    $B^0\rightarrow \pi^0\eta_1$&&& $B^0_s\rightarrow \pi^0\eta_1$\\
    $B^0\rightarrow \eta_8\eta_1$&&& $B^0_s\rightarrow\eta_8\eta_1$\\
    \hline
    \vspace{-0.3cm}&&&\\
    $B^0_s\rightarrow\eta_1{\bar K^0}$&&&$B^0\rightarrow\eta_1K^0$\\
    \\
    \multicolumn{4}{c}{$\boldsymbol{1\otimes 1}$}\\
    \hline
    \multicolumn{2}{c|}{$\Delta S = 0$} &
    \multicolumn{2}{c}{$\Delta S = 1$}\\
    \hline\hline
    \vspace{-0.3cm}&&&\\
    $B^0\rightarrow \eta_1\eta_1$&&& $B^0_s\rightarrow \eta_1\eta_1$\\
\end{tabular}
\end{center}
    \caption{$\Delta S=0$ and $\Delta S=1$ $B \to PP$ decays, including $\eta_8$ and/or $\eta_1$ in the final state.}
    \label{tab:BPPdecays}
\end{table}

Consider first the $\boldsymbol{8\otimes 8}$ $B \to PP$ decays. The amplitudes for these decays are related under SU(3)$_F$: the Wigner-Eckart theorem can be used to express all the amplitudes in terms of a small number of SU(3)$_F$ reduced matrix elements. The observables associated with all of these decays (branching ratios, direct and indirect CP asymmetries) can then be written as functions of these RMEs and a fit to the data can be performed. This will tell us how well the SU(3)$_F$ limit of the SM agrees with the data. 

Instead of using RMEs as the theoretical parameters, an alternative method uses topological quark diagrams \cite{Gronau:1994rj, Gronau:1995hn}. As noted in the Introduction, diagrams have certain potential advantages over RMEs for the purposes of this analysis. But the equivalence of diagrams and RMEs must be demonstrated. 

In this section, for each of the three $B \to PP$ categories, $\boldsymbol{8\otimes 8}$, $\boldsymbol{8\otimes 1}$ and $\boldsymbol{1\otimes 1}$, we describe the amplitudes for all of the decays in terms of RMEs and diagrams, and demonstrate the equivalence of the two approaches.

\subsection{\boldmath RMEs}\label{sec:RMEs}

The analysis involves the application of the Wigner-Eckart theorem to the $B \to PP$ amplitudes within SU(3)$_F$ symmetry. The first step is to determine how each of the pieces of the matrix element $\langle PP | H_W | B \rangle$ -- the initial state, the weak Hamiltonian, and the final state -- transforms under SU(3)$_F$.

Charmless $B \to PP$ decays involve the quark-level transitions ${\bar b} \to {\bar u} u {\bar q}$ and ${\bar b}
\to {\bar q}$, $q = d, s$. These decays are governed by the weak Hamiltonian \cite{Buchalla:1995vs}
\beq
H_W = \frac{G_F}{\sqrt{2}} \sum_{q=d,s}\left(\lambda_u^{(q)} \sum_{i=1}^{2} c_i Q_i^{(q)} 
- \lambda_t^{(q)} \sum_{i=3}^{10} c_i Q_i^{(q)} \right) ~.
\label{eq:HW}
\eeq
Here $\lambda_{q'}^{(q)} \equiv V_{q'b}^* V_{q'q}$, $q=d,s$, $q' = u,c,t$, where the $V_{ij}$ are elements of the Cabibbo-Kobayashi-Maskawa (CKM) matrix. The $Q_1^{(q)}$-$Q_2^{(q)}$, $Q_3^{(q)}$-$Q_6^{(q)}$ and $Q_7^{(q)}$-$Q_{10}^{(q)}$ represent tree, gluonic penguin and electroweak penguin operators, respectively. The $c_i$ ($i = 1$-10) are Wilson coefficients. 

The quarks $(u,d,s)$ form a triplet $\boldsymbol{3}$ under SU(3)$_F$ ($(-{\bar u},{\bar d},{\bar s})$ form an antitriplet $\boldsymbol{3^*}$). Given the transitions ${\bar b} \to {\bar u} u {\bar q}$ and ${\bar b} \to {\bar q}$, it is straightforward to show that $H_W$ transforms as a ${\bf 3^{*(a)}}$, ${\bf 3^{*(s)}}$, ${\bf 6}$, or ${\bf 15^*}$ of
SU(3)$_F$. 

The individual operators of $H_W$ contribute differently to the various $B \to PP$ amplitudes. For this reason, it is necessary to determine how these operators transform under SU(3)$_F$. The operators proportional to $\lambda_u^{(q)}$ are
\bea
Q_1^{(q)} &=& ({\bar b}u)_{V-A}({\bar u}q)_{V-A} ~, \nn\\
Q_2^{(q)} &=& ({\bar b}q)_{V-A}({\bar u}u)_{V-A} ~.
\label{Q1,2operators}
\eea
They transform as $\boldsymbol{3^*\otimes 3 \otimes 3^* = 3^{*(a)} \oplus 3^{*(s)} \oplus 6 \oplus 15^*}$.
The four gluonic penguin operators are defined as follows \cite{Gronau:1998fn}:
\begin{widetext}
\begin{equation}
\begin{split}
    Q_{3}^{(q)}&=({\bar b}q)_{V-A}(({\bar u}u)_{V-A}+({\bar d}d)_{V-A}+({\bar s}s)_{V-A}) ~, \\
    Q_{4}^{(q)}&=({\bar b}u)_{V-A}({\bar u}q)_{V-A}+({\bar b}d)_{V-A}({\bar d}q)_{V-A}+({\bar b}s)_{V-A}({\bar s}q)_{V-A} ~, \\
    Q_{5}^{(q)}&=({\bar b}q)_{V-A}(({\bar u}u)_{V+A}+({\bar d}d)_{V+A}+({\bar s}s)_{V+A}) ~, \\
    Q_{6}^{(q)}&=({\bar b}u)_{V-A}({\bar u}q)_{V+A}+({\bar b}d)_{V-A}({\bar d}q)_{V+A}+({\bar b}s)_{V-A}({\bar s}q)_{V+A} ~.
    \label{eq:gluon_oper}
\end{split}
\end{equation}
These operators transform as two triplets of SU(3)$_F$, $\boldsymbol{3^{*(a)}}$ and $\boldsymbol{3^{*(s)}}$. 
Finally, the four electroweak penguin operators are
\begin{equation}
\begin{split}
    Q_{7}^{(q)}&=\frac{3}{2}\left[({\bar b}q)_{V-A}\left(\frac{2}{3}({\bar u}u)_{V+A}-\frac{1}{3}({\bar d}d)_{V+A}-\frac{1}{3}({\bar s}s)_{V+A}\right)\right] ~, \\
    Q_{8}^{(q)}&=\frac{3}{2}\left[\frac{2}{3}({\bar b}u)_{V-A}({\bar u}q)_{V+A}-\frac{1}{3}({\bar b}d)_{V-A}({\bar d}q)_{V+A}-\frac{1}{3}({\bar b}s)_{V-A}({\bar s}q)_{V+A}\right] ~, \\
    Q_{9}^{(q)}&=\frac{3}{2}\left[({\bar b}q)_{V-A}\left(\frac{2}{3}({\bar u}u)_{V-A}-\frac{1}{3}({\bar d}d)_{V-A}-\frac{1}{3}({\bar s}s)_{V-A}\right)\right] ~, \\
    Q_{10}^{(q)}&=\frac{3}{2}\left[\frac{2}{3}({\bar b}u)_{V-A}({\bar u}q)_{V-A}-\frac{1}{3}({\bar b}d)_{V-A}({\bar d}q)_{V-A}-\frac{1}{3}({\bar b}s)_{V-A}({\bar s}q)_{V-A}\right] ~.
\end{split}
\label{eq:EWP_oper}
\end{equation}
\end{widetext}
These operators couple to the electric charge. The product of the quarks with this added coupling gives four operators: $\boldsymbol{3^{*(a)}},\boldsymbol{3^{*(s)}},\boldsymbol{6}$ and $\boldsymbol{15^*}$. 

Although these electroweak penguin operators, with coefficient $\lambda_t^{(q)}$, involve the same SU(3)$_F$ representations as $Q_1^{(q)}$ and $Q_2^{(q)}$ [Eq.(\ref{Q1,2operators})], with coefficient $\lambda_u^{(q)}$, the linear combination of $\boldsymbol{3^{*(a)}}$ and $\boldsymbol{3^{*(s)}}$ is different. This allows us to redefine these two representations in a more convenient way. From here on, the linear combination with coefficient $\lambda_u^{(q)}$ ($\lambda_t^{(q)}$) will be referred to as $\boldsymbol{3^*_1}$ ($\boldsymbol{3^*_2}$). 

When we write $H_W$ as a function of the SU(3)$_F$ representations, the terms containing ${\bf 3^*_1}$, ${\bf 3^*_2}$, ${\bf 6}$, and ${\bf 15^*}$ can have particular $c_i$s as coefficients. Note that the values of $c_7$ and $c_8$ are much smaller than those of $c_9$ and $c_{10}$ \cite{Buchalla:1995vs}, so they are neglected in our study. The $c_i$ that appear in terms containing the four representations are as follows:
\begin{equation}
\begin{split}
    \boldsymbol{3^*_1} & : c_1,c_2, \\
\boldsymbol{3^*_2} & : c_3,c_4,c_5,c_6,c_9,c_{10}, \\
    \boldsymbol{6} & : c_1,c_2,c_9,c_{10}, \\
    \boldsymbol{15^*} & : c_1,c_2,c_9,c_{10}.
\end{split}
\end{equation}

The initial state $B = \{B^+ = {\bar b} u, B^0 = {\bar b} d, B^0_s = {\bar b} s \}$ is a $\boldsymbol{3}$ of SU(3)$_F$.

Turning to the $PP$ final states, as noted above, they can be separated into three categories, $\boldsymbol{8\otimes8}$, $\boldsymbol{8\otimes1}$, and $\boldsymbol{1\otimes1}$. Consider first $\boldsymbol{8\otimes8}$. All particles in the octet are considered to be identical under SU(3)$_F$, so that the $\boldsymbol{8\otimes8}$ final state must be symmetrized under the exchange of the two particles: $(\boldsymbol{8\otimes8})_S=\boldsymbol{1\oplus8\oplus27}$. 

Combining all the pieces of the matrix elements, charmless $B \to PP$ decays with an $(\boldsymbol{8\otimes8})_S$ final state
are described by seven RMEs. When the amplitudes are written as functions of the RMEs, they appear with certain CKM factors as coefficients. Including these CKM factors, the seven RMEs are 
\bea
\lambda_u^{(q)} &:& A_1 = \langle {\bf 1} || {\bf 3^*_1} || {\bf 3} \rangle_{({\bf 8}\otimes{\bf 8})_S} ~,~\nn\\ && A_8 = \langle {\bf 8} || {\bf 3^*_1} || {\bf 3} \rangle_{({\bf 8}\otimes{\bf 8})_S} ~,\nn \\
\lambda_t^{(q)} &:& B_1 = \langle {\bf 1} || {\bf 3^*_2} || {\bf 3} \rangle_{({\bf 8}\otimes{\bf 8})_S} ~,~\nn\\ && B_8 = \langle {\bf 8} || {\bf 3^*_2} || {\bf 3} \rangle_{({\bf 8}\otimes{\bf 8})_S} ~,\nn \\
\lambda_u^{(q)}~{\&}~\lambda_t^{(q)} &:& R_8 = \langle {\bf 8} || {\bf 6} || {\bf 3} \rangle_{({\bf 8}\otimes{\bf 8})_S} ~,~ \nn\\
&& P_8 = \langle {\bf 8} || {\bf 15^*} || {\bf 3} \rangle_{({\bf 8}\otimes{\bf 8})_S}~,~\nn\\
&&P_{27} = \langle {\bf 27} || {\bf 15^*} || {\bf 3} \rangle_{({\bf 8}\otimes{\bf 8})_S} ~.
\label{8x8RMEs}
\eea
We must stress that this result is rigorous from a group-theoretical point of view. If SU(3)$_F$ is unbroken, all amplitudes with an $(\boldsymbol{8\otimes8})_S$ final state are expressed in terms of these seven complex parameters. And since the amplitudes are linear combinations of these parameters, using a different basis does not change the number of independent parameters. For example, as we will see in the following subsection, an alternative description of the amplitudes can be given in terms of diagrams. But there are a total of 14 diagrams. This means that there must exist seven relationships among these diagrams, in order to reduce the total number of independent parameters to seven, the same number as the RMEs above.

We can now use the Wigner-Eckart theorem to express the amplitudes for all $B \to PP$ decays with an $(\boldsymbol{8\otimes8})_S$ final state in terms of the RMEs in Eq.~(\ref{8x8RMEs}). These amplitude decompositions are given in Table \ref{tab:8x8RME} in Appendix \ref{app:RMEs}.

A similar exercise can be carried out for the $\boldsymbol{8\otimes1}$ final state, which transforms simply as an $\boldsymbol{8}$. There are four RMEs:
\bea
\label{8x1RMEs}
\lambda_u^{(q)} &:& C_8 = \langle {\bf 8} || {\bf 3^*_1} || {\bf 3} \rangle_{{\bf 8}\otimes{\bf 1}} ~, \nn \\
\lambda_t^{(q)} &:& D_8 = \langle {\bf 8} || {\bf 3^*_2} || {\bf 3} \rangle_{{\bf 8}\otimes{\bf 1}} ~, \\
\lambda_u^{(q)}~{\&}~\lambda_t^{(q)} &:& L_8 = \langle {\bf 8} || {\bf 6} || {\bf 3} \rangle_{{\bf 8}\otimes{\bf 1}} ~,~ 
M_8 = \langle {\bf 8} || {\bf 15^*} || {\bf 3} \rangle_{{\bf 8}\otimes{\bf 1}} ~. \nn
\eea
The amplitude decompositions for the $\boldsymbol{8\otimes1}$ decays in terms of the above RMEs are given in Table \ref{tab:8x1RME} in Appendix \ref{app:RMEs}.

Finally, the $\boldsymbol{1\otimes1}$ final state transforms as a $\boldsymbol{1}$ and has two RMEs:
\bea
\lambda_u^{(q)} &:& C_1 = \langle {\bf 1} || {\bf 3^*_1} || {\bf 3} \rangle_{{\bf 1}\otimes{\bf 1}} ~,~\nn \\
\lambda_t^{(q)} &:& D_1 = \langle {\bf 1} || {\bf 3^*_2} || {\bf 3} \rangle_{{\bf 1}\otimes{\bf 1}} ~.
\label{1x1RMEs}
\eea
The amplitude decompositions for the $\boldsymbol{1\otimes1}$ decays can be found in Table \ref{tab:1x1RME} in Appendix \ref{app:RMEs}.

\subsection{\boldmath Diagrams}
\label{Sec:diagrams}

It is also possible to express the $B \to PP$ amplitudes in terms of topological quark diagrams. In
Ref.~\cite{Gronau:1994rj}, the following six diagrams were introduced: $T$ (tree), $C$ (colour-suppressed tree), $P$ (penguin), $A$ (annihilation), $E$ (exchange) and $PA$ (penguin annihilation). 

$T$, $C$, $A$ and $E$ are all proportional to $\lambda_u^{(q)}$. The diagrams $P$ and $PA$ each contain loops with internal $u$, $c$ and $t$ quarks, proportional to $\lambda_u^{(q)}$, $\lambda_c^{(q)}$ and $\lambda_t^{(q)}$, respectively. The unitarity of the CKM matrix implies that $\lambda_u^{(q)} + \lambda_c^{(q)} + \lambda_t^{(q)} = 0$. Using this relation
to eliminate the $c$-quark pieces of $P$ and $PA$, we obtain combinations of diagrams proportional to $\lambda_u^{(q)}$ ($P_{uc}$ and $PA_{uc}$) or $\lambda_t^{(q)}$ ($P_{tc}$ and $PA_{tc}$).

Two other diagrams were added in Ref.~\cite{Gronau:1995hn}: $P_{EW}$ (electroweak penguin [EWP]) and $P_{EW}^C$ (colour-suppressed electroweak penguin). It was later pointed out in Ref.~\cite{Gronau:1998fn} that, in fact, each diagram proportional to $\lambda_u^{(q)}$ has its own EWP counterpart. Four additional EWP diagrams must therefore be introduced: $P_{EW}^A$, $P_{EW}^E$, $P_{EW}^{P_u}$ and $P_{EW}^{PA_{u}}$. All EWP diagrams are proportional to $\lambda_t^{(q)}$. Figures of all six EWP diagrams can be found in Appendix \ref{app:EWPs}.

These four additional EWPs were not included in the analysis of Ref.~\cite{Berthiaume:2023kmp}. As they are expected to be small, it is unlikely that their inclusion would change the results very much. Still, we will check this. 

There are, therefore, a total of 14 diagrams (or combinations of diagrams) that contribute to our $B \to PP$ decays. Six are proportional to $\lambda_u^{(q)}$ and eight are proportional to $\lambda_t^{(q)}$. In Tables {\ref{tab:8x8DIAG}}, {\ref{tab:8x1DIAG}} and {\ref{tab:1x1DIAG}} in Appendix \ref{app:diags}, we decompose all the $\Delta S = 0$ and $\Delta S=1$ $B \to PP$ amplitudes in terms of these diagrams for the $(\boldsymbol{8\otimes 8})_S$, $\boldsymbol{8\otimes 1}$ and $\boldsymbol{1\otimes 1}$ final states, respectively. Note that we denote the diagrams in these three decay categories by the same symbols ($T$, $C$, etc.). However, it must be understood that the diagrams in the three categories are different.

\subsection{\boldmath Relations between RMEs and diagrams}

We have now expressed all the $B \to PP$ decay amplitudes in terms of two distinct bases, RMEs and diagrams. Since the physics must be basis-independent, the two decompositions must be equivalent, i.e., there must be relations between the sets of RMEs and diagrams. 

However, there is a complication. As detailed above, there are far more diagrams than there are RMEs. This means that not all the diagrams are independent: they can be written in terms of fewer effective diagrams. There are then relations between the RMEs and these effective diagrams. In this subsection, we derive all of these relations.

\subsubsection{The $\lambda^{(q)}_u$ sector}

We have seen that the $B \to PP$ amplitudes with the $(\boldsymbol{8\otimes8})_S$ final state are described by five RMEs, $A_1$, $A_8$, $R_8$, $P_8$ and $P_{27}$ [Eq.~(\ref{8x8RMEs})]. However, there are six diagrams, $T$, $C$, $P_{uc}$, $A$, $E$ and $PA_{uc}$. These diagrams, therefore, cannot be independent.

The amplitudes for these decays are decomposed in terms of diagrams in Table {\ref{tab:8x8DIAG}} in Appendix \ref{app:diags}. For the six diagrams proportional to $\lambda_u^{(q)}$, only five linear combinations appear in the amplitudes. (This fact was pointed out in Ref.~\cite{Gronau:1994rj}.) In light of this, we have chosen to eliminate the $E$ diagram, defining five effective diagrams as follows: 
\begin{equation}
\begin{split}
    \widetilde{T} &\equiv T + E~,\\
    \widetilde{C} &\equiv C - E~,\\
    \widetilde{P_{uc}} &\equiv P_{uc} - E~,\\
    \widetilde{A} &\equiv A + E ~,\\
    \widetilde{PA_{uc}} &\equiv PA_{uc} + E ~.
\end{split}
\label{lambda_uDIAGS}
\end{equation}
The relations between the RMEs and these effective diagrams are as follows (see also Ref.~\cite{Berthiaume:2023kmp}):
\begin{equation}
\begin{split}
    A_1 &= \frac{1}{2\sqrt{3}}\left(-3\tT + \tC - 8\tPuc -12\tPAuc\right)\\
    A_8 &= \frac{1}{8}\sqrt{\frac{5}{3}}\left(-3\tT +\tC -8\tPuc -3\tA\right)\\
    R_8 &= \frac{\sqrt{5}}{4}\left(\tT-\tC-\tA \right)\\
    P_8 &= \frac{1}{8\sqrt{3}}\left(\tT +\tC +5\tA\right)\\
    P_{27} &= -\frac{1}{2\sqrt{3}}\left(\tT +\tC\right)
\end{split}
\label{eq:8x8_RMEtoDIAG}
\end{equation}

A similar analysis can be carried out for the decays involving $\boldsymbol{8\otimes 1}$ $PP$ final states. There are three RMEs proportional to $\lambda_u^{(q)}$, $C_8$, $L_8$ and $M_8$ [Eq.~(\ref{8x1RMEs})]. But there are once again six diagrams, $T$, $C$, $P_{uc}$, $A$, $E$ and $PA_{uc}$. 
(We remind the reader that, although these diagrams are denoted by the same symbols as the $(\boldsymbol{8\otimes 8})_S$ diagrams, they are not the same.)
Consulting Table {\ref{tab:8x8DIAG}} in Appendix \ref{app:diags}, which contains the decomposition of the $\boldsymbol{8\otimes 1}$ amplitudes in terms of diagrams, we find that only three independent effective diagrams are involved:
\bea
&&\tT \equiv T+2A~,~~\tC \equiv C+2E~,~~\tPuc \equiv P_{uc}-E ~.
\eea
The relations between the RMEs and these effective diagrams are 
\begin{equation}
\begin{split}
    C_8 &= \frac{1}{8\sqrt{3}}\left(3\tT +7\tC +16\tPuc \right) ~, \\
    L_8 &= \frac{1}{4}\left(\tT - \tC\right) ~, \\
    M_8 &= -\frac{\sqrt{5}}{8\sqrt{3}}\left(\tT+\tC\right) ~. \\
\end{split}
\label{eq:8x1_RMEtoDIAG}
\end{equation}

Finally, for the $\boldsymbol{1\otimes 1}$ final state, there is one RME proportional to $\lambda_u^{(q)}$, $C_1$ [Eq.~(\ref{1x1RMEs})]. 
From Table {\ref{tab:1x1DIAG} in Appendix \ref{app:diags}, we see there is only one effective diagram, $\widetilde{C} \equiv C + P_{uc} + E + 6PA_{uc}$, leading to a trivial RME-diagram relation, 
\begin{equation}
    C_1 = \sqrt{\frac{2}{3}}\widetilde{C} ~.
\end{equation}

\subsubsection{The $\lambda^{(q)}_t$ sector}

From Eq.~(\ref{8x8RMEs}), we see that the $B \to PP$ amplitudes with the $(\boldsymbol{8\otimes8})_S$ final state are again described by five RMEs, $B_1$, $B_8$, $R_8$, $P_8$ and $P_{27}$. But in the $\lambda^{(q)}_t$ sector, there are eight diagrams, $P_{tc}$, $PA_{tc}$, $P_{EW}$, $P_{EW}^C$, $P_{EW}^A$, $P_{EW}^E$, $P_{EW}^{P_u}$ and $P_{EW}^{PA_{u}}$.

As was done in the $\lambda^{(q)}_u$ sector, we can reduce the number of diagrams by diagram redefinitions. First, we note that $P_{EW}^E$ can be eliminated by redefining the other EWP diagrams: 
\begin{equation}
\begin{split}
    \widetilde{P_{EW}^T} \equiv P_{EW}^T + P_{EW}^E ~&,~\widetilde{P_{EW}^C} \equiv P_{EW}^C - P_{EW}^E ~, \\
    \widetilde{P_{EW}^{P_u}} \equiv P_{EW}^{P_u}  - P_{EW}^E ~&,~\widetilde{P_{EW}^A} \equiv P_{EW}^{A} + P_{EW}^E ~,\\
    \widetilde{P_{EW}^{PA_{u}}} \equiv~ &P_{EW}^{PA_{u}} + P_{EW}^E ~.
\end{split}
\end{equation}
Second, we can make two further redefinitions:
\begin{equation}
\begin{split}
    \widetilde{P_{tc}} &\equiv P_{tc} - \frac{1}{3}\widetilde{P^{P_u}_{EW}} ~, \\
    \widetilde{PA_{tc}} &\equiv PA_{tc} - \frac{1}{3}\widetilde{PA_{EW}^{PA_u}} ~.
\end{split}
\label{EWPredef}
\end{equation}
With these five effective diagrams, the amplitudes in Table {\ref{tab:8x8DIAG}} in Appendix \ref{app:diags} are unchanged.

For the terms proportional to $\lambda_t^{(q)}$, we therefore have five effective diagrams and five RMEs. Writing the diagrams as a sum of RMEs, we have
\begin{equation}
\begin{split}
    \widetilde{P_{tc}} &= -\sqrt{\frac{3}{5}}B_8 +\frac{1}{\sqrt{5}}R_8 +\sqrt{3}P_8 ~, \\
    \widetilde{PA_{tc}} &= -\frac{1}{2\sqrt{3}}B_1 +\frac{2}{\sqrt{15}}B_8 -\frac{2\sqrt{3}}{5}P_8 -\frac{\sqrt{3}}{10}P_{27} ~, \\
    \widetilde{P_{EW}^{T}} &= \frac{6}{\sqrt{5}}R_8 + \frac{12\sqrt{3}}{5}P_8 - \frac{12\sqrt{3}}{5}P_{27} ~, \\
    \widetilde{P_{EW}^{C}} &= -\frac{6}{\sqrt{5}}R_8 - \frac{12\sqrt{3}}{5}P_8 - \frac{18\sqrt{3}}{5}P_{27} ~, \\
    \widetilde{P_{EW}^A} &= \frac{24\sqrt{3}}{5}P_8 + \frac{6\sqrt{3}}{5}P_{27} ~.
\end{split}
\label{eq:8x8_EWPtoRME}
\end{equation}
Note that the three EWP diagrams are functions of the RMEs $R_8$, $P_8$, and $P_{27}$. But in Eq.~(\ref{eq:8x8_RMEtoDIAG}), these RMEs are given as functions of $\tT$, $\tC$ and $\tA$. This is a key point: because of this, there are three additional relations relating these EWP diagrams in the $\lambda_t^{(q)}$ sector to three ``tree'' diagrams in the $\lambda_u^{(q)}$ sector. In finding the correspondence between the RMEs in the two sectors, it is necessary to properly take into account the Wilson coefficients as well as a factor of $-1/2$. When this is done, the EWP-tree relations are
\bea
\label{eq:EWPtree8x8}
    \widetilde{P^T_{EW}} &=& -\frac{3}{4}\left[\frac{\Sigma_9}{\Sigma_1}(\tT+\tC+\tA)+\frac{\Delta_9}{\Delta_1}(\tT-\tC-\tA)\right] ~, \nn\\
    \widetilde{P^C_{EW}} &=& -\frac{3}{4}\left[\frac{\Sigma_9}{\Sigma_1}(\tT+\tC-\tA)-\frac{\Delta_9}{\Delta_1}(\tT-\tC-\tA)\right] ~, \nn\\
    \widetilde{P^A_{EW}} &=& -\frac{3}{2}\frac{\Sigma_9}{\Sigma_1}\tA ~,
\eea
where $\Sigma_1 = c_1+c_2$, $\Sigma_9= c_9+c_{10}$, $\Delta_1=c_1-c_2$, and $\Delta_9 = c_9-c_{10}$. If one makes the approximation that ${\Sigma_9}/{\Sigma_1} = {\Delta_9}/{\Delta_1}$, which holds to better than 3\%, then the EWP-tree relations simplify to 
\beq
\widetilde{P^T_{EW}} = -\frac32 \kappa \tT ~,~~ \widetilde{P^C_{EW}} = -\frac32 \kappa \tC 
~,~~ \widetilde{P^A_{EW}} = -\frac32 \kappa \tA ~,
\eeq
where $\kappa \simeq -1.123\alpha$ \cite{Gronau:1998fn}.

For the $\boldsymbol{8\otimes 1}$ final state, there are six diagrams in the $\lambda_t^{(q)}$ sector: $P_{tc}$, $P_{EW}$, $P_{EW}^C$, $P_{EW}^A$, $P_{EW}^E$ and $P_{EW}^{P_u}$. For the EWP diagrams, we find that there are only three independent effective diagrams:
\begin{equation}
\begin{split}
    \widetilde{P_{EW}^T} \equiv P_{EW}^T + 2P_{EW}^A ~&,~ \widetilde{P_{EW}^C} \equiv P_{EW}^C + 2P_{EW}^E ~, \\
    \widetilde{P_{EW}^{P_{u}}} \equiv &~ P_{EW}^{P_u} - P_{EW}^E ~.
\end{split}
\end{equation}
As was done in Eq.~(\ref{EWPredef}) for the $(\boldsymbol{8}\otimes \boldsymbol{8})_S$ case, one further redefinition can be made:
\begin{equation}
    \widetilde{P_{tc}} \equiv P_{tc} -\frac{1}{3} \widetilde{P_{EW}^{P_{u}}} ~.
\end{equation}

For the $\boldsymbol{8\otimes1}$ final states, we therefore find that there are three effective diagrams proportional to $\lambda_t^{(q)}$ and three RMEs. Once again, writing the diagrams as a sum of RMEs, we have
\begin{equation}
\begin{split}
    \widetilde{P_{tc}} &= \frac{\sqrt{3}}{2}D_8  -\frac{1}{2}L_8 -\frac{\sqrt{15}}{2}M_8 ~, \\
    \widetilde{P_{EW}^{T}} &= 6L_8-12\sqrt{\frac{3}{5}}M_8 ~, \\
    \widetilde{P_{EW}^C} &= -6L_8-12\sqrt{\frac{3}{5}}M_8 ~.
\end{split}
\label{eq:8x1_EWPtoRME}
\end{equation}
Similar to the $(\boldsymbol{8\otimes 8})_S$ case, the two EWP diagrams are functions of the RMEs $L_8$ and $M_8$, but these RMEs are given as functions of $\tT$ and $\tC$ in Eq.~(\ref{eq:8x1_RMEtoDIAG}). This leads to two EWP-tree relations:
\begin{equation}
\begin{split}
    \widetilde{P_{EW}^T} &= -\frac{3}{4}\left[\frac{\Sigma_9}{\Sigma_1}(\tT+\tC) + \frac{\Delta_9}{\Delta_1}(\tT-\tC)\right] ~, \\
    \widetilde{P_{EW}^C} &=  -\frac{3}{4}\left[\frac{\Sigma_9}{\Sigma_1}(\tT+\tC) - \frac{\Delta_9}{\Delta_1}(\tT-\tC)\right] ~. \\
\end{split}
\label{eq:EWPtree8x1}
\end{equation}
We note that both relations for $P_{EW}^T$ and $P_{EW}^C$ from Eqs. (\ref{eq:EWPtree8x8}) and (\ref{eq:EWPtree8x1}) are the same, considering that $A$ has been absorbed for $\boldsymbol{8\otimes 1}$.

Finally, there is one effective diagram proportional to $\lambda^{(q)}_t$ for the $\boldsymbol{1\otimes 1}$ final state:
\beq
    \widetilde{P_{tc}} \equiv P_{tc} + 6PA_{tc} -\frac{1}{3}P_{EW}^C -\frac{1}{3}P_{EW}^E - \frac{1}{3}P_{EW}^{P_u} ~.
\eeq
There is only one RME, $D_1$, leading to the RME-diagram relation
\begin{equation}
    \widetilde{P_{tc}} = \sqrt{\frac{3}{2}}D_1 ~.
\end{equation}

\subsubsection{Recap}

For the $B \to PP$ decays with the $(\boldsymbol{8\otimes8})_S$ final state, the amplitudes are described by seven RMEs. Two are proportional to $\lambda^{(q)}_u$, two are proportional to $\lambda^{(q)}_t$, and three can be proportional to $\lambda^{(q)}_u$ or $\lambda^{(q)}_t$. Equivalently, the amplitudes are described by five effective diagrams proportional to $\lambda^{(q)}_u$ and two effective diagrams proportional to $\lambda^{(q)}_t$. There are also three EWP diagrams proportional to $\lambda^{(q)}_t$, but these can be written in terms of three diagrams proportional to $\lambda^{(q)}_u$ using the EWP-tree relations of Eq.~(\ref{eq:EWPtree8x8}).

For the $\boldsymbol{8\otimes1}$ final states, the amplitudes are described by four RMEs. One is proportional to $\lambda^{(q)}_u$, one is proportional to $\lambda^{(q)}_t$, and two can be proportional to $\lambda^{(q)}_u$ or $\lambda^{(q)}_t$. The amplitudes are equivalently described in terms of four effective diagrams. Three are proportional to $\lambda^{(q)}_u$ and one is proportional to $\lambda^{(q)}_t$. There are also two EWP diagrams proportional to $\lambda^{(q)}_t$, but these can be written in terms of two diagrams proportional to $\lambda^{(q)}_u$ using the EWP-tree relations of Eq.~(\ref{eq:EWPtree8x1}).

Finally, for the $\boldsymbol{1\otimes1}$ final state, there is one RME and one effective diagram proportional to $\lambda^{(q)}_u$, and one RME and one effective diagram proportional to $\lambda^{(q)}_t$. The RME-diagram relations are trivial.
There is no EWP-tree relation.

\section{Fits to the Data}

\subsection{Observables}

There are a total of 34 charmless $B \to PP$ decays. Of these, 28 have been measured. In these measurements, three types of observables are obtained: CP-averaged branching ratios (${\cal B}_{CP}$), direct CP asymmetries ($A_{CP}$), and indirect CP asymmetries ($S_{CP}$). These are defined as follows: 
\beq
{\cal B}_{CP} = F_{\rm PS} \, (|A|^2 + |{\bar A}|^2) ~, 
\eeq
where
\beq
    F_{PS}=\frac{\sqrt{m_B^2-(m_{P_1}+m_{P_2})^2}\sqrt{m_B^2-(m_{P_1}-m_{P_2})^2}S}{32\pi m^3_B\Gamma_B} ~, \nn
\eeq
\bea
    & A_{CP} =\frac{|{\bar A}|^2-|A|^2}{|{\bar A}|^2+|A|^2} ~, & \\
    & S_{CP} = \eta_{CP} \times 2\text{Im} \, \left(\frac{q}{p}\frac{{\bar A}A^*}{|{\bar A}|^2 + |A|^2}\right) ~. &
    \label{eq:cp_asymmetries}
\eea
In the above, $A$ and ${\bar A}$ are the amplitudes for $B \to PP$ and its
CP-conjugate process, respectively, $S$ is a statistical factor
related to the presence of identical particles in the final state, and $q/p = \exp(-2
i \phi_M)$, where $\phi_M$ is the weak phase of $B_q^0$-${\bar B}_q^0$
mixing. In the definition of $S_{CP}$, the factor $\eta_{CP}$ takes the value $\pm1$ depending on whether or not the final state is a CP eigenstate.

A complete list of the presently-measured observables, along with their experimental values, can be found in Table \ref{tab:exp_data} in Appendix \ref{app:data}. (Note that quite a few of these have changed since the fits were done in Ref.~\cite{Berthiaume:2023kmp}.) For the direct CP asymmetry, some experiments present the result for $C_{CP} = -A_{CP}$. In this Table, we have added the appropriate minus signs so that all results are for $A_{CP}$.

\subsection{\boldmath $\eta$ and $\eta'$}
\label{eta,eta'}

We note that, when the $B \to PP$ amplitudes are expressed in terms of RMEs or diagrams (Tables \ref{tab:8x8RME}-\ref{tab:1x1DIAG}), the particles that appear in the final states are eigenstates of SU(3)$_F$. In particular, one finds $\eta_8$ and $\eta_1$, respectively part of the octet and a singlet under SU(3)$_F$. However, the physical particles that are measured -- $\eta$ and $\eta'$ -- are different. 

The point is that $\eta$ and $\eta'$ are mixtures of $\eta_8$ and $\eta_1$. We write
\begin{equation}
\begin{split}
    \eta &= \eta_8 \cos \theta_\eta - \eta_1\sin \theta_\eta ~, \\
    \eta' &= \eta_8\sin\theta_\eta + \eta_1 \cos\theta_\eta ~,
\end{split}
\label{thetaetadef}
\end{equation}
where
\begin{equation}
    \eta_8 \equiv \frac{(2s{\bar s}-u{\bar u} -d{\bar d})}{\sqrt{6}}, ~~~ \eta_1 \equiv \frac{(u{\bar u} + d{\bar d} + s{\bar s})}{\sqrt{3}} ~.
\end{equation}
It is common practice to take $\theta_\eta = {\rm arcsin}(1/3) \approx 19.5^\circ$, so that \cite{Kawarabayashi:1980dp, Gilman:1987ax, Chau:1990ay, Dighe:1995gq}
\begin{equation}
    \eta = \frac{(s{\bar s} - u{\bar u} - d{\bar d})}{\sqrt{3}}, ~~~ \eta' =\frac{(2s{\bar s} + u{\bar u} +d{\bar d})}{\sqrt{6}} ~.
\end{equation}
In our fits, we almost always take the above value for $\theta_\eta$; it is kept as a free variable in one fit.

\subsection{\boldmath Fits without $\eta$, $\eta'$}

As noted above, the four EWP diagrams $P_{EW}^A$, $P_{EW}^E$, $P_{EW}^{P_u}$ and $P_{EW}^{PA_{u}}$ were not included in the analysis of Ref.~\cite{Berthiaume:2023kmp}. Our first task is to redo this analysis, including these diagrams, to see how the previous results change. 

In Ref.~\cite{Berthiaume:2023kmp}, the focus was on $B \to PP$ decays in which the $P$ is a pion or a kaon. There are a total of 16 such decays, 8 with $\Delta S=0$ and 8 with $\Delta S=1$. They can be found in the table of $\boldsymbol{8\otimes8}$ decays given in Sec.~\ref{Sec:BPPamps}. 

Consider first the $\Delta S=0$ decays. The decomposition of the amplitudes in terms of diagrams is given in Table \ref{tab:8x8DIAG} in Appendix \ref{app:diags}. Taking into account the analysis in Sec.~\ref{Sec:BPPamps}B, these amplitudes are functions of 7 effective diagrams. 5 are proportional to $\lambda_u^{(q)}$: $\widetilde{T}$, $\widetilde{C}$, $\widetilde{P_{uc}}$, $\widetilde{A}$ and $\widetilde{PA}_{uc}$ [Eq.~(\ref{lambda_uDIAGS})], and 2 are proportional to $\lambda_t^{(q)}$: $\widetilde{P_{tc}}$ and $\widetilde{PA_{tc}}$ [Eq.~(\ref{EWPredef})]. $P_{EW}^E$, $P_{EW}^{P_u}$ and $P_{EW}^{PA_{u}}$ have been absorbed into these effective diagrams, leaving three EWPs: $P_{EW}^T$, $P_{EW}^C$ and $P_{EW}^A$. But these are related to $\widetilde{T}$, $\widetilde{C}$ and $\widetilde{A}$ by the EWP-tree relations of Eq.~(\ref{eq:EWPtree8x8}). 

These 7 diagrams correspond to 13 unknown theoretical parameters (7 magnitudes, 6 relative strong phases). The amplitudes also depend on other quantities: the weak phases $\gamma$, $\beta$ (in $B^0$-${\bar B}^0$ mixing) and $\phi_s$ (in $B_s^0$-${\bar B}_s^0$ mixing), as well as the CKM matrix elements involved in $\lambda_{u,t}^{(q)}$. Values for all of these quantities are taken from the Particle Data Group (PDG) \cite{ParticleDataGroup:2024cfk}.

There are 15 measured observables for these decays (see Table \ref{tab:exp_data} in Appendix \ref{app:data}). With more observables than unknown parameters, a fit can be performed. We define the $\chi^2$ function as follows:
\begin{equation}
    \chi^2 = \sum_i \left(\frac{O^{\text{th}}_i - O^{\text{exp}}_i}{\Delta O^{\text{exp}}_i}\right)^2 ~.
\end{equation}
Here the sum is over all the observables $O_i$. We vary the theoretical parameters, and for each set of parameters, we compute the theoretical values of the observables, $O^{\rm th}_i$. By comparing them with the experimental values, $O^{\rm exp}_i \pm \Delta O^{\rm exp}_i$, we can find the set of parameters that best reproduces the data, i.e., that minimizes the $\chi^2$.

The fit is excellent: we find $\chi_{\rm min}^2/{\rm d.o.f.} = 1.06/2$, for a $p$-value of 0.59. The conclusion is that the $\Delta S=0$ data can be explained by the SM$_{\rm{SU(3)}_F}$. 

For $\Delta S=1$ decays, there are again 13 unknown theoretical parameters, along with 15 measured observables (see Table \ref{tab:exp_data} in Appendix \ref{app:data}). Performing the fit, we find that the fit is slightly worse than for $\Delta S=0$ decays, but still good: $\chi_{\rm min}^2/{\rm d.o.f.} = 1.55/2$, for a $p$-value of 0.46.

Within the SU(3)$_F$ symmetry, the diagrams in $\Delta S=0$ and $\Delta S=1$ decays are the same, and we can perform a combined fit including all the data. But here we find that the fit is very poor: the best fit has $\chi_{\rm min}^2/{\rm d.o.f.} = 43.2/17$, for a $p$-value of $4.5 \times 10^{-4}$. That is, the data disagree with the SM$_{\rm{SU(3)}_F}$ at the level of $3.5\sigma$.

We therefore find that, even when the four EWP diagrams $P_{EW}^A$, $P_{EW}^E$, $P_{EW}^{P_u}$ and $P_{EW}^{PA_{u}}$ are included in the analysis, the results of all three fits are very similar to those of Ref.~\cite{Berthiaume:2023kmp}. 

We also confirm that the data exhibit very large SU(3)$_F$ breaking. In Table \ref{tab:fitresults8x8} in Appendix \ref{app:res}, we show the best-fit values of the theoretical parameters for each fit.  From these, we compute
the ratios of the magnitudes of the $\Delta S = 1$ diagrams ($D'$) and the corresponding $\Delta S = 0$ diagrams ($D$). For the three largest diagrams, these ratios are given in the second column of Table \ref{tab:diags_ratios}. In the SU(3)$_F$ limit, we expect $|D'/D| = 1$. However, focusing on $|\tT'/\tT|$ and $|\tC'/\tC|$, the average of their central values is 11, which corresponds to greater than 1000\% SU(3)$_F$ breaking. This is much larger than the $\sim 20\%$ expected in the SM (from $f_K/f_\pi$), and perhaps suggests the presence of new physics \footnote{We note in passing that there have been several papers studying the effects of new physics on the $\btopik$ puzzle \cite{Crivellin:2019isj, Calibbi:2019lvs, Bhattacharya:2021shk, Bhattacharya:2024clv, Datta:2024zrl}. And new-physics explanations of the anomalies in Ref.~\cite{Amhis:2022hpm} have been searched for in Ref.~\cite{Grossman:2024amc}. However, to date there has been no comprehensive study of new-physics contributions to all $B\to PP$ decays.}.

\begin{table}[H]
    \centering
    \begin{tabular}{c | c c c} \hline
     & \multicolumn{3}{c}{Type of fit} \\ \cline{2-4}
        Ratios &~ no $\eta/\eta'$~ &~ at most one $\eta/\eta'$~&~ all the data~\\
        \hline\hline
        $|\tT'/\tT|$ & $13.1\pm2.1$ & $13.1\pm1.9$ & $12.5\pm1.7$ \\
        $|\tC'/\tC|$ & $8.9\pm2.3$ & $8.9\pm1.5$ & $8.6\pm1.5$ \\
        $|\tPuc'/\tPuc|$ & $23\pm54$ & $23\pm6$ & $22\pm5$\\
        \hline
    \end{tabular}
    \caption{Ratios $|D'/D|$ of the three largest diagrams when comparing the fits for $\Delta S = 0$ ($D$ diagrams) and $\Delta S = 1$ ($D'$ diagrams) decays.}
    \label{tab:diags_ratios}
\end{table}

The bottom line is that, even with the inclusion of the additional EWP diagrams, and with new measurements of some observables in Table \ref{tab:exp_data} in Appendix \ref{app:data}, the conclusions of Ref.~\cite{Berthiaume:2023kmp} still hold.

\subsection{Fits with $\mathbf{\eta}$, $\mathbf{\eta'}$}

We now expand the fits to include observables associated with decays in which one or both of the final-state mesons is an $\eta$ or $\eta'$, taking the $\eta$-$\eta'$ mixing angle of Eq.~(\ref{thetaetadef}) to be $\theta_\eta =19.5^\circ$. These decays are of all three types, $(\boldsymbol{8\otimes8})_S$, $\boldsymbol{8\otimes1}$ and $\boldsymbol{1\otimes1}$. Since the diagrams for these three decay categories are distinct, we have a total of $7 + 4 + 2 = 13$ independent diagrams, corresponding to 25 free parameters (13 magnitudes and 12 relative strong phases). Not all diagrams contribute to all the decays. Decays with only pions and kaons in the final state involve only $(\boldsymbol{8\otimes 8})_S$ diagrams. Decays which have only one $\eta$ or $\eta'$ in the final state have contributions from $(\boldsymbol{8\otimes 8})_S$ and $\boldsymbol{8\otimes 1}$ diagrams. Finally, only $\eta\eta$, $\eta\eta'$ and $\eta'\eta'$ final states involve $\boldsymbol{1\otimes1}$ diagrams.

We have already performed the $(\boldsymbol{8\otimes8})_S$ fits. There are 30 observables in the decays
involving final states made up of only pions and kaons: 15 are in $\Delta S = 0$ decays and 15 are in $\Delta S = 1$ decays. There are 7 diagrams, corresponding to 13 parameters, which means that a fit combining all the data under the assumption of SU(3)$_F$ can be performed. As detailed in the previous subsection, we find a very poor combined fit: the data disagree with the SM$_{\rm{SU(3)}_F}$ at the level of $3.5\sigma$. 

We now add $\boldsymbol{8\otimes1}$ decays, in which the final state consists of a $\pi$ or $K$ along with an $\eta$ or $\eta'$. There are 12 such decays (6 each of $\Delta S = 0$ and $\Delta S = 1$), leading to 14 additional observables (6 $\Delta S = 0$ and 8 $\Delta S = 1$). There are now a total of 11 diagrams, or 21 parameters. If we assume SU(3)$_F$ and include all the data in the fit, we once again get a very poor fit: we find $\chi^2_{\text{min}}/\text{d.o.f}= 55.6/23$, which corresponds to a $p$-value of $1.6\times 10^{-4}$, or a $3.8\sigma$ discrepancy with the SM$_{\rm{SU(3)}_F}$.

Finally, we also include the $\boldsymbol{1\otimes1}$ decays involving two $\eta/\eta'$ mesons. There are now a total of 13 diagrams, or 25 parameters. With 49 observables (23 $\Delta S = 0$ and 26 $\Delta S=1$), we can perform a fit using all the data under the assumption of SU(3)$_F$. The fit is even poorer than the previous combined fits: we find $\chi^2_{\text{min}}/\text{d.o.f}= 60.8/24$, corresponding to a $p$-value of $4.9\times10^{-5}$, i.e., a $4.1\sigma$ discrepancy with the SM$_{\rm{SU(3)}_F}$. 

We note in passing that, when this fit is redone allowing $\theta_\eta$ to be a free parameter, the best-fit value is very close to $19.5^\circ$.

We therefore find that, when we add decays involving one or two $\eta$ or $\eta'$ mesons, the combined fits become increasingly poor. What is the reason for these very poor fits? When we restricted the fits to $(\boldsymbol{8\otimes8})_S$ decays, we were able to perform individual $\Delta S = 0$ and $\Delta S = 1$ fits. As detailed in the previous subsection, both individual fits were good. Comparing the results of the individual $\Delta S = 0$ and $\Delta S = 1$ fits, we determined that the very poor combined fit was due to the fact that SU(3)$_F$ is badly broken (see the second column of Table \ref{tab:diags_ratios}).  

When we add $\boldsymbol{8\otimes1}$ decays, the comparison of the individual $\Delta S = 0$ and $\Delta S = 1$ fits is more complicated. In the $\Delta S = 1$ sector, we have 23 observables and 21 parameters, so a fit can be performed. We find a very good fit: $\chi^2_{\text{min}}/{\text{d.o.f}}=1.55/2$, corresponding to a $p$-value of 0.46. This said, the number of d.o.f is perhaps a bit misleading, since not all parameters contribute to all observables. In $\Delta S = 0$ decays, there are 21 observables and 21 parameters, which suggests that a fit can be done. However, note that 8 of these parameters correspond to the 4 diagrams in the $\boldsymbol{8\otimes1}$ decays, but there are only 6 observables. If the equations were linear, there would be an infinite number of solutions. However, since the equations are nonlinear, we are not guaranteed that there is even one solution. And indeed, when we perform a fit (when there are more parameters than observables, this is known as ``overfitting''), we find no exact solution: $\chi^2_{\text{min}}=1.06$.

When $\boldsymbol{1\otimes1}$ decays are added, both of the $\Delta S = 0$ and $\Delta S = 1$ individual fits are overfitted. The fit results are respectively $\chi^2_{\text{min}}=1.06$ and $\chi^2_{\text{min}}=2.44$. Note that, for the $\Delta S = 0$ decays, we find the same $\chi^2_{\text{min}}$ as was found for the $\boldsymbol{8\otimes1}$ decays. Considering that there are only two additional observables, but four new parameters, the fit is able to perfectly reproduce the $\boldsymbol{1\otimes1}$ data, even if the equations are nonlinear.

It therefore appears that, when decays involving final-state $\eta$/$\eta'$ mesons are included, the comparison of the results of the individual $\Delta S = 0$ and $\Delta S = 1$ fits  is not easy because several of these are overfitted. Fortunately, the fact is that {\it all} the fits include decays whose final states comprise only pions and kaons. This means that, for the $(\boldsymbol{8\otimes8})_S$ diagrams, there are \emph{always} more observables than parameters. This subset of diagrams is, therefore, never overfitted. 

The upshot is that, even in fits including $\boldsymbol{8\otimes1}$ and $\boldsymbol{1\otimes1}$ decays, we can focus only on the best-fit values of the $(\boldsymbol{8\otimes8})_S$ diagrams. In particular, using the best-fit values of the theoretical parameters (see Table \ref{tab:fitresults} in Appendix \ref{app:res}), 
we can compute the ratios of the magnitudes of several $\Delta S = 1$ diagrams ($D'$) and the corresponding $\Delta S = 0$ diagrams ($D$). The results can be found in the third and fourth columns of Table \ref{tab:diags_ratios}.
We see that, even when the $\eta/\eta'$ data is added, these ratios are still $O(10)$ or larger. We conclude that $\sim 1000\%$ $SU(3)_F$ breaking is still required to properly explain the data, and this is responsible for the very poor fits, i.e., the sizeable discrepancies with the SM$_{\rm{SU(3)}_F}$.

We note that theoretical estimates of the decay constant of the $\eta_8$ give $f_{\eta_8} \sim 115$ MeV \cite{Bali:2021qem}, similar to $f_\pi$. We therefore expect that, even when final states with $\eta$ and $\eta'$ are included in the analysis, the SM prediction for SU(3)$_F$ breaking is still $\approx 20$\%. This is much smaller than the $\sim 1000$\% SU(3)$_F$ breaking observed in the data.

\subsection{Bayesian analysis}

In the previous subsections, the fits were all done using the \textit{Minuit} package \cite{James:1975dr} to find the minimum $\chi^2$. We have also checked our results by performing a Bayesian analysis with the \textit{dynesty} package \cite{10.1093/mnras/staa278}. 

One advantage of the Bayesian analysis is that it gives us a better picture of the allowed ranges of the parameters, as well as their mutual correlations. In Fig.~\ref{fig:dynesty_fit}, we present examples of  ``corner plots'' that can be drawn with the results of the Bayesian regression algorithm, focusing specifically on the $\tT$ and $\tC$ diagrams of the $\boldsymbol{8\otimes 8}$ sector. We show the allowed ranges for ${\rm Re}(\tT)$, ${\rm Re}(\tC)$ and ${\rm Im}(\tC)$ (we take ${\rm Im}(\tT) = 0$), as well as the correlations among them. This is done for (a) the fit with only $\Delta S=0$ data, (b) the fit with only $\Delta S=1$ data, and (c) the fit with all the data.

These plots come in two types. Each two-dimensional (off-diagonal) scatter plot is a topological map showing the correlation between two parameters. Here, a dense cluster represents a high probability for the parameters' values. When the distribution is circular, the parameters are not correlated, but when the distribution is elongated and oriented diagonally, there is some degree of correlation. There are also plots on the diagonal, showing the probabilistic distribution of the preferred values of the individual parameters.

We note the following. First, on all the scatter plots, we overlay the results of the \textit{Minuit} fits. We see that the \textit{Minuit} values of $\chi^2_{\rm min}$ lie within the $\sim 1\sigma$ ranges of the \textit{dynesty} values, showing that the two methods are in good agreement. Second, looking at the diagonal plots, we see that the theoretical parameters are all reasonably Gaussian:
\begin{figure}[!htbp]
\begin{center}
    \includegraphics[width=0.33\textwidth]{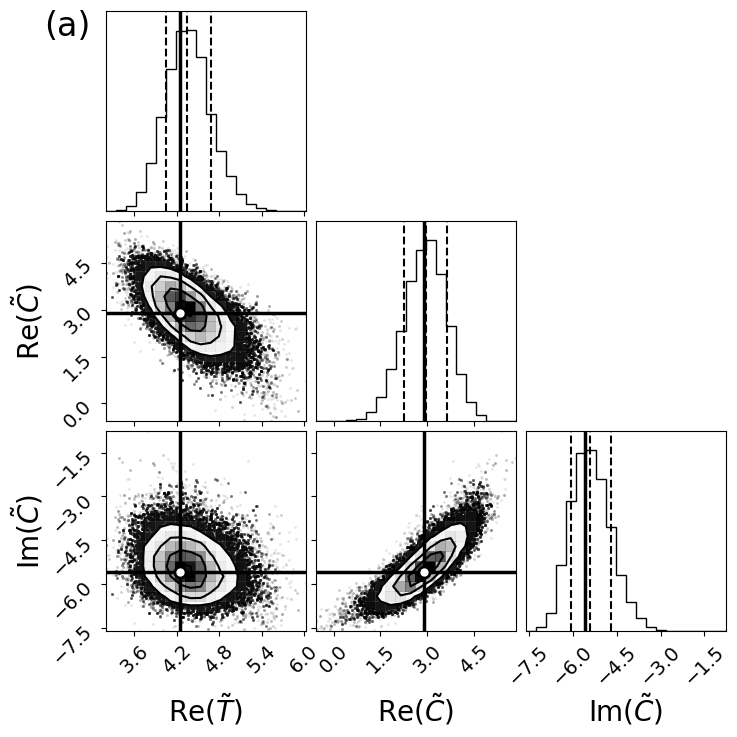} \vskip 5pt
    \includegraphics[width=0.33\textwidth]{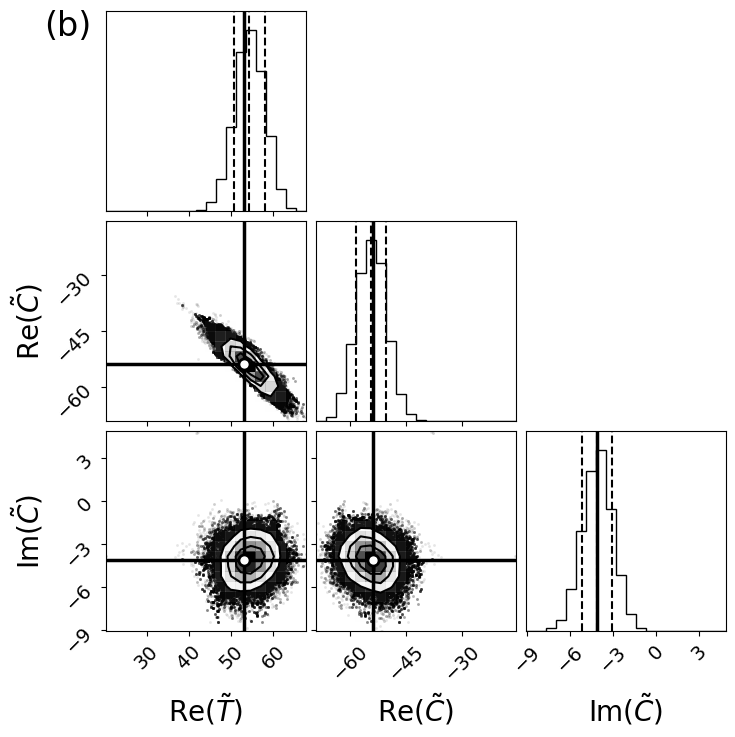} \vskip 5pt
    \includegraphics[width=0.33\textwidth]{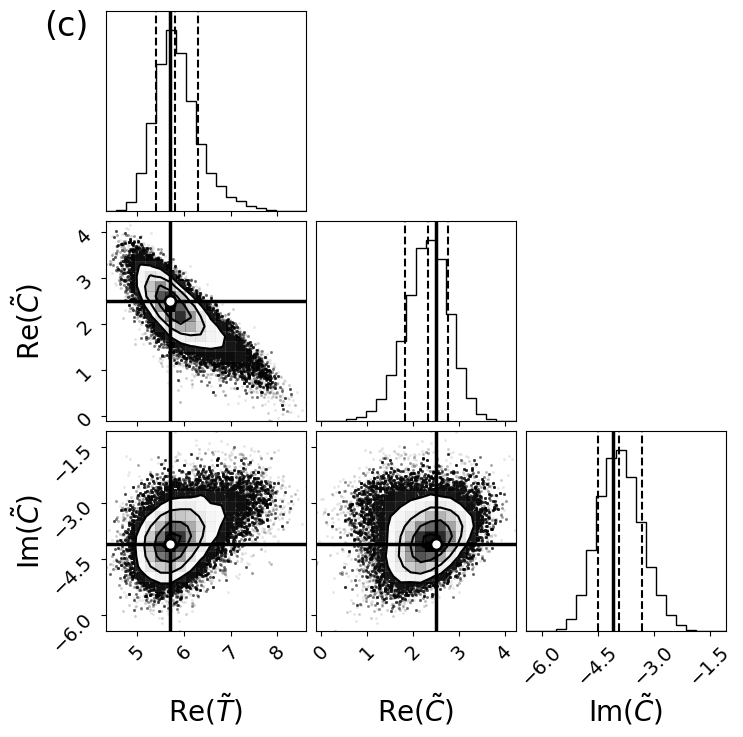} \vskip -5pt
\end{center}
\caption{Parameter distributions for fits using the \textit{dynesty} package \cite{10.1093/mnras/staa278} for (a) the fit with only $\Delta S=0$ data, (b) the fit with only $\Delta S=1$ data, and (c) the fit with all the data. The dashed lines represent the mean and standard deviation obtained using \textit{dynesty}. The thick lines represent the central values of the fits done with \textit{Minuit}. We see that both methods are in good agreement.}
\label{fig:dynesty_fit}
\end{figure}
\bea
\Delta S=0~{\rm fit} &:& ~~~~~ {\rm Re}(\tT) = 4.4 \pm 0.3 ~, \nn\\
&& {\rm Re}(\tC) = 2.9 \pm 0.7 ~,~ {\rm Im}(\tC) = -5.4^{+0.7}_{-0.6} ~, \nn\\
\Delta S=1~{\rm fit} &:& ~~~~~ {\rm Re}(\tT) = 54.3^{+3.6}_{-3.5} ~,~ \nn\\
&& {\rm Re}(\tC) = -54.4^{+3.9}_{-4.1}  ~,~ {\rm Im}(\tC) = -4.1^{+1.0}_{-1.1} ~, \nn\\
{\rm full~fit} &:& ~~~~~ {\rm Re}(\tT) = 5.8^{+0.5}_{-0.4} ~, \\
&& {\rm Re}(\tC) = 2.3^{+0.4}_{-0.5} ~,~ {\rm Im}(\tC) = -3.9 \pm 0.6 ~. \nn
\eea

These best-fit central values agree well with those found using \textit{Minuit}.
Finally, averaging the asymmetric errors in quadrature, we find the following ratios for the magnitudes of diagrams:
\beq
|\tT'/\tT| = 12.5 \pm 1.7 ~~,~~~~ |\tC'/\tC| = 8.8 \pm 1.7 ~.
\eeq
These are very similar to the ratios in Table \ref{tab:diags_ratios}, and show that the Bayesian fits exhibit the same 1000\% SU(3)$_F$ breaking as was found using the minimization technique.

We therefore see that the results of the fits using \textit{Minuit} are largely reproduced when \textit{dynesty} is used. Although we show the corner plots for only the first three parameters, this trend is followed by all the parameters.

\subsection{Comparison with the literature}
\label{Comparison_literature}

Our results disagree with some other results in the literature. For example, in Ref.~\cite{Hsiao:2015iiu} (2016), it is found that the $(\boldsymbol{8\otimes8})_S$ data can be well-explained within the SM$_{\rm{SU(3)}_F}$. This analysis is extended to include decays with $\eta$/$\eta'$ in the final state in
Ref.~\cite{Huber:2021cgk} (2022), and once again a good global fit is found. Finally, in a recent update of the fit to the $(\boldsymbol{8\otimes8})_S$ data \cite{BurgosMarcos:2025xja}, it is found that the fit is not satisfactory: $\chi_{\rm min}^2/{\rm d.o.f.} = 32.3/15$, for a $p$-value of $5.8 \times 10^{-3}$ (recall that our $p$-value was $4.5 \times 10^{-4}$). But they did not find a large breaking of SU(3)$_F$ -- they determined that a good fit to the data could be found with the addition of factorizable SU(3)$_F$-breaking effects.

The difference between these analyses and ours comes down to the number of free parameters. The above three analyses follow the formalism of Ref.~\cite{He:2018php}, in which there are ten complex parameters, for a total of 19 unknowns. But we have only seven diagrams, or 13 unknowns. With more free parameters, it is not surprising that better fits are obtained in the above analyses.

The source of this disagreement is in the counting of RMEs (or diagrams). In Ref.~\cite{He:2018php}, it is argued that there are five RMEs proportional to $\lambda_u$ and five proportional to $\lambda_t$. While this is true, we have shown in Eq.~(\ref{8x8RMEs}) that not all ten of these RMEs are independent: three of the RMEs proportional to $\lambda_t$ are identical to three proportional to $\lambda_u$. 

Taking ten independent RMEs is equivalent to ignoring the EWP-tree relations of Eq.~(\ref{eq:EWPtree8x8}). Indeed, when we redo the combined fit without imposing the EWP-tree relations, we obtain a $\chi_{\rm min}^2$ similar to that of Ref.~\cite{BurgosMarcos:2025xja}. For this fit, in Table \ref{tab:no_EWP_tree_relations} we show the best-fit values of the magnitudes of the diagrams involved in the EWP-tree relations. It is clear that these relations are badly broken: the best-fit values of the EWPs are 30 times or more larger than their values predicted by the EWP-tree relations. This is also a reflection of 1000\% SU(3)$_F$ breaking.

\begin{table}[H]
\begin{center}
\begin{tabular}{|c|c|c|c|} \hline
Fit combining & $|\tT|$ & $|\tC|$ & $|\tA|$\\ \cline{2-4}
$\Delta S = 0$ and  $\Delta S = 1$ & $5.0\pm0.3$ & $4.5\pm0.4$ & $0.00\pm0.12$ \\ \cline{2-4}
decays, assuming& $|P_{EW}^T|_{\rm fit}$ & $|P_{EW}^C|_{\rm fit}$ & $|P_{EW}^A|_{\rm fit}$  \\ \cline{2-4}
SU(3)$_F$ & $2.33\pm0.02$ & $2.37\pm0.02$ & $2.20\pm0.12$  \\ \hline\hline         
Values expected from & $|P_{EW}^T|_{\rm exp}$ & $|P_{EW}^C|_{\rm exp}$ & $|P_{EW}^A|_{\rm exp}$  \\ \cline{2-4}
EWP-tree relations & 0.067 & 0.059 & 0.00 \\ \hline

Ratios fit/expected & $|P_{EW}^T|$ & $|P_{EW}^C|$ & $|P_{EW}^A|$  \\ \cline{2-4}
& 34.8 & 40.2 & $\infty$ \\ \hline

\end{tabular}
\end{center}
    \caption{Best-fit values of the magnitudes of some diagrams for the combined fit in which the EWP-tree relations are not used. The ratios of the best-fit values of the EWP diagrams and those predicted by the EWP-tree relations are also shown.}
    \label{tab:no_EWP_tree_relations}
\end{table}

\subsection{Theoretical Input}

Up to now, the analysis has been completely rigorous, group-theoretically -- no theoretical input has been used. In this section, we examine the effect of adding certain well-motivated theoretical assumptions to the fits.

\subsubsection{$|C/T| = 0.2$}

Naively, the ratio $|C/T|$ is expected to equal 1/3, simply by counting colors. This is supported by theoretical calculations: within QCD factorization, this ratio is computed for $B \to \pi K$ decays and $|C/T| \simeq 0.2$ is found \cite{Beneke:2001ev, Bell:2007tv, Bell:2009nk, Beneke:2009ek, Bell:2015koa}. 

In light of this, we fix $|\tilde{C}/\tilde{T}| = 0.2$ for the $(\boldsymbol{8\otimes 8})_S$ diagrams and redo the fits. We now find that the fit combining the $\Delta S=0$ and $\Delta S=1$ observables under the assumption of SU(3)$_F$ has $\chi^2_{\text{min}}/\text{d.o.f.} = 74.0/25$, corresponding to a $p$-value of $9.6\times10^{-7}$, or a $4.9\sigma$ discrepancy with the SM$_{\rm{SU(3)}_F}$.

We can also add the same restriction to the $\boldsymbol{8\otimes1}$ parameters. The effect is very similar: we now find $\chi^2_{\text{min}}/\text{d.o.f.} = 74.5/26$, for a $p$-value of $1.4\times10^{-6}$, i.e., a $4.8\sigma$ discrepancy with the SM$_{\rm{SU(3)}_F}$. The small difference between the results of these two fits comes from the large experimental errors associated with the decays with one or more $\eta/\eta'$ in the final state. 

\subsubsection{$A = 0$}

$A$, $E$ and $PA$ all require the $b$ quark to interact with the spectator quark. For this reason, these diagrams are expected to be much smaller than the others \cite{Gronau:1994rj}, and are often simply neglected. In our analysis, we have kept these diagrams. A linear combination of $E$ and $PA$ is constrained to be small by measurements, which suggests that the individual $E$ and $PA$ diagrams are also small. On the other hand, there is no $B \to PP$ decay that is dominated by $A$, so that $A$ is not strongly constrained by the data.

We note, however, that the decays $B^+ \to D^+ K^0$ and $B^+ \to D_s^+ \phi$ are pure-$A$ decays, and are found to be small: $B(B^+ \to D^+ K^0) < 2 \times 10^{-6}$ and $B(B^+ \to D_s^+ \phi) < 2 \times 10^{-7}$ \cite{ParticleDataGroup:2024cfk}. While the $A$ diagram in $B \to DP$ decays is unrelated by SU(3)$_F$ to the $A$ in $B \to PP$ decays, it is interesting to see that at least one $A$ diagram is constrained by data to be small.

In light of this, we look at what happens to the fits when we set the value of $A(A')$ to be 0. The individual $\Delta S = 0$ and $\Delta S = 1$ fits are not much affected by this constraint: they have $\chi^2_{\text{min}}/{\text{d.o.f.}} = 1.09/4$ ($p$-value $=0.90$ and $\chi^2_{\text{min}}/{\text{d.o.f.}} = 3.09/4$ ($p$-value $=0.54$), respectively. On the other hand, the combined fit experiences a larger change:  $\chi^2_{\text{min}}/{\text{d.o.f.}} = 68.2/19$, for a $p$-value of $1.83\times10^{-7}$, or a $5.2\sigma$ discrepancy with the SM$_{\rm{SU(3)}_F}$.

\section{Conclusions}

Recently, $B\to PP$ decays ($B = \{B^0, B^+, B_s^0\}$, $P = \{ \pi, K \}$) were analyzed under the assumption of flavor SU(3) symmetry (SU(3)$_F$) \cite{Berthiaume:2023kmp}. It was found that, although the individual fits to $\Delta S=0$ or $\Delta S=1$ decays are good, the combined fit is very poor: there is a $3.6\sigma$ disagreement with the SU(3)$_F$ limit of the standard model (SM$_{\rm{SU(3)}_F}$). This discrepancy can be removed by adding SU(3)$_F$-breaking effects, but 1000\% SU(3)$_F$ breaking is required.

In this paper, we extend this analysis to include decays in which there is an $\eta$ and/or $\eta'$ meson in the final state. The three $\pi$s and four $K$s are members of the meson octet of SU(3)$_F$, ${\bf 8}$. The eighth member is $\eta_8$. There is also a singlet, $\eta_1 = {\bf 1}$. The $\eta$ and $\eta'$ are mixtures of $\eta_8$ and $\eta_1$ (we take the mixing angle to be $\theta_\eta = 19.5^\circ$). We therefore consider three categories of decays, those in which the final state transforms as $\boldsymbol{8\otimes8}$, $\boldsymbol{8\otimes1}$, and $\boldsymbol{1\otimes1}$.

For all three categories, we write the amplitudes in terms of both SU(3)$_F$ reduced matrix elements (RMEs) and topological diagrams, and demonstrate the equivalence of the two descriptions. We find that there are seven independent RMEs in $\boldsymbol{8\otimes8}$ decays, four in $\boldsymbol{8\otimes1}$ decays, and two in $\boldsymbol{1\otimes1}$ decays. 

We first update the fits using decays in which the final state consists of only pions or kaons (the $\boldsymbol{8\otimes8}$ sector). We find that the individual fits using only $\Delta S=0$ or $\Delta S=1$ decays are good, but the combined fit is very poor: we find $\chi_{\rm min}^2/{\rm d.o.f.} = 43.2/17$, for a $p$-value of $4.5 \times 10^{-4}$, i.e., a disagreement with the SM$_{\rm{SU(3)}_F}$ at the level of $3.5\sigma$. These results are very similar to those of Ref.~\cite{Berthiaume:2023kmp}.

We then add decays that contain a single $\eta$ or $\eta'$ ($\boldsymbol{8\otimes1}$ decays). Now the combined fit has $\chi^2_{\text{min}}/\text{d.o.f}= 55.6/23$, which corresponds to a $p$-value of $1.6\times 10^{-4}$, or a $3.8\sigma$ discrepancy with the SM$_{\rm{SU(3)}_F}$. Finally, we include the $\boldsymbol{1\otimes1}$ decays involving two $\eta/\eta'$ mesons. We find $\chi^2_{\text{min}}/\text{d.o.f}= 60.8/24$, corresponding to a $p$-value of $4.9\times10^{-5}$, i.e., a $4.1\sigma$ discrepancy with the SM$_{\rm{SU(3)}_F}$. To summarize, when decays with $\eta$ and/or $\eta'$ in the final state are added to the analysis of Ref.~\cite{Berthiaume:2023kmp}, the disagreement with the SM$_{\rm{SU(3)}_F}$ increases.

All three fits include the decays in which the final state consists of only pions or kaons (the $\boldsymbol{8\otimes8}$ sector). For these decays, we can also do individual fits involving only $\Delta S=0$ and $\Delta S=1$ decays. For all three types of fit, we find that, for the largest diagrams, the best-fit values in $\Delta S=1$ decays are $\sim 10$ times larger than the best-fit values in $\Delta S=0$ decays. However, these should be equal in the SU(3)$_F$ limit. We therefore deduce that $\sim 1000\%$ SU(3)$_F$ breaking is present in the data, and that this is responsible for the very poor fits.

Finally, the above analysis is rigorous, group-theoretically -- no theoretical input has been used. However, if we assume that the annihilation diagram $A=0$ -- it is expected to be much smaller than the largest diagrams -- we find that the discrepancy with the SM$_{\rm{SU(3)}_F}$ grows to $4.4\sigma$. And if instead we fix the ratio of diagrams $|C/T|   = 0.2$, which is the value found within QCD factorization, there is a $4.9\sigma$ disagreement with the SM$_{\rm{SU(3)}_F}$.

\bigskip\bigskip
\noindent
{\bf Acknowledgements:} We thank Gilberto Tetlalmatzi-Xolocotzi for helpful conversations. This work was financially supported by the National Science Foundation, Grant No.\ PHY-2310627 (BB, LH, CM), by NSERC of Canada  (MB, AJ, DL), and by FRQNT, Scholarship No.\ 363240 (MB).

\bibliography{b2ppetas_final}
\bibliographystyle{apsrev4-2}

\appendix

\onecolumngrid
 
\newpage

\section{$B \to PP$ amplitudes: decomposition in terms of RMEs}
\label{app:RMEs}

In $B \to PP$ decays, the final-state pseudoscalar $P$ can be a member of the SU(3)$_F$ octet ${\bf 8} = \{\pi^\pm, \pi^0, K^\pm, K^0, {\bar K}^0, \eta_8 \}$ or the singlet ${\bf 1} = \eta_1$. (The physical $\eta$ and $\eta'$ mesons are linear combinations of $\eta_8$ and $\eta_1$, see Sec.~\ref{eta,eta'}.) The $PP$ final state therefore comes in three categories: $(\boldsymbol{8\otimes 8})_S$, $\boldsymbol{8\otimes1}$, or $\boldsymbol{1\otimes1}$. Each category has its own set of RMEs; these are defined in Eqs.~(\ref{8x8RMEs}), (\ref{8x1RMEs}) and (\ref{1x1RMEs}), respectively. All $B \to PP$ decay amplitudes can be decomposed in terms of RMEs. For the three categories of decays, this decomposition is shown below in Tables \ref{tab:8x8RME}, \ref{tab:8x1RME} and \ref{tab:1x1RME}, respectively.

\begin{table}[H]
\begin{center}
    \begin{tabular}{|l|l|c|c|c|c|c||c|c|c|c|c|}
         \hline
         \multicolumn{2}{|c|}{\multirow{3}{*}{Decays}} & \multicolumn{5}{c||}{$\lambda_u^{(q)}$} & \multicolumn{5}{c|}{$\lambda_t^{(q)}$}\\
         \cline{3-12}
         \multicolumn{2}{|l|}{}& \multicolumn{5}{c||}{$c_1,c_2$} & \multicolumn{2}{c|}{$c_3$, $c_4$, $c_5$, $c_6$, $c_9$, $c_{10}$} & \multicolumn{3}{c|}{$c_9$, $c_{10}$}\\
         \cline{3-12}
         \multicolumn{2}{|l|}{}& $A_1$ & $A_8$ & $R_8$ & $P_8$ & $P_{27}$ & $B_1$ & $B_8$ & $R_8$ & $P_8$ & $P_{27}$\\
         \hline\hline
         \multirow{12}{*}{$\Delta S=0$}
         &$B^+\ra \Kb K^+$ & 0 & $-\frac{\sqrt{3}}{\sqrt{5}}$ & $-\frac{1}{\sqrt{5}}$ & $\frac{3\sqrt{3}}{5}$ & $\frac{2\sqrt{3}}{5}$ & 0 & $-\frac{\sqrt{3}}{\sqrt{5}}$ & $\frac{3}{\sqrt{5}}$ & $\frac{9\sqrt{3}}{5}$ & $\frac{6\sqrt{3}}{5}$\\
         &$B^+\ra \pi^0\pi^+$ & 0 & 0 & 0 & 0 & $\sqrt{6}$ & 0 & 0 & 0 & 0 & $3\sqrt{6}$\\ 
         &$B^+\ra \eta_8 \pi^+$ & 0 & $\frac{\sqrt{2}}{\sqrt{5}}$ & $\frac{\sqrt{2}}{\sqrt{15}}$ & $-\frac{3\sqrt{2}}{5}$ & $\frac{3\sqrt{2}}{5}$&0 & $\frac{\sqrt{2}}{\sqrt{5}}$ & $-\frac{\sqrt{6}}{\sqrt{5}}$ & $-\frac{9\sqrt{2}}{5}$ & $\frac{9\sqrt{2}}{5}$\\
         \cline{2-12}
         &$B^0\ra K^0\Kb$& $-\frac{1}{2\sqrt{3}}$ & $-\frac{1}{\sqrt{15}}$ & $-\frac{1}{\sqrt{5}}$ & $-\frac{3\sqrt{3}}{5}$ & $\frac{\sqrt{3}}{10}$ & $-\frac{1}{2\sqrt{3}}$ & $-\frac{1}{\sqrt{15}}$ & $\frac{3}{\sqrt{5}}$ & $-\frac{9\sqrt{3}}{5}$ & $\frac{3\sqrt{3}}{10}$\\
         &$B^0\ra \pi^+\pi^-$ & $\frac{1}{2\sqrt{3}}$ & $\frac{1}{\sqrt{15}}$ & $-\frac{1}{\sqrt{5}}$ & $-\frac{\sqrt{3}}{5}$ & $\frac{7\sqrt{3}}{10}$ & $\frac{1}{2\sqrt{3}}$ & $\frac{1}{\sqrt{15}}$ & $\frac{3}{\sqrt{5}}$ & $-\frac{3\sqrt{3}}{5}$ & $\frac{21\sqrt{3}}{10}$\\
         &$B^0\ra K^-K^+$ & $\frac{1}{2\sqrt{3}}$ & $-\frac{2}{\sqrt{15}}$ & 0 & $-\frac{2\sqrt{3}}{5}$ & $-\frac{\sqrt{3}}{10}$ & $\frac{1}{2\sqrt{3}}$ & $-\frac{2}{\sqrt{15}}$ & 0 & $-\frac{6\sqrt{3}}{5}$ & $-\frac{3\sqrt{3}}{10}$\\
         &$B^0\ra \pi^0\pi^0$ & $-\frac{1}{2\sqrt{6}}$ & $-\frac{1}{\sqrt{30}}$ & $\frac{1}{\sqrt{10}}$ & $\frac{\sqrt{3}}{5\sqrt{2}}$ & $\frac{13\sqrt{3}}{10\sqrt{2}}$ & $-\frac{1}{2\sqrt{6}}$ & $-\frac{1}{\sqrt{30}}$ & $-\frac{3}{\sqrt{10}}$ & $\frac{3\sqrt{3}}{5\sqrt{2}}$ & $\frac{39\sqrt{3}}{10\sqrt{2}}$\\
         &$B^0\ra \pi^0\eta_8$ & 0 & $\frac{1}{\sqrt{5}}$ & $\frac{1}{\sqrt{15}}$ & 1 & 0 & 0 & $\frac{1}{\sqrt{5}}$ & $-\frac{\sqrt{3}}{\sqrt{5}}$ & $3$ & 0\\
         &$B^0 \ra \eta_8\eta_8$ & $-\frac{1}{2\sqrt{6}}$ & $\frac{1}{\sqrt{30}}$ & $-\frac{1}{\sqrt{10}}$ & $-\frac{\sqrt{3}}{5\sqrt{2}}$ & $-\frac{3\sqrt{3}}{10\sqrt{2}}$ & $-\frac{1}{2\sqrt{6}}$ & $\frac{1}{\sqrt{30}}$ & $\frac{3}{\sqrt{10}}$ & $-\frac{3\sqrt{3}}{5\sqrt{2}}$ & $-\frac{9\sqrt{3}}{10\sqrt{2}}$\\
         \cline{2-12}
         &$B^0_s \ra \pi^+K^-$ & 0 & $\frac{\sqrt{3}}{\sqrt{5}}$ & $-\frac{1}{\sqrt{5}}$ & $\frac{\sqrt{3}}{5}$ & $\frac{4\sqrt{3}}{5}$ & 0 & $\frac{\sqrt{3}}{\sqrt{5}}$ & $\frac{3}{\sqrt{5}}$ & $\frac{3\sqrt{3}}{5}$ & $\frac{12\sqrt{3}}{5}$\\
         &$B^0_s \ra \pi^0\Kb$ & 0 & $-\frac{\sqrt{3}}{\sqrt{10}}$ & $\frac{1}{\sqrt{10}}$ & $-\frac{\sqrt{3}}{5\sqrt{2}}$ & $\frac{3\sqrt{6}}{5}$ & 0 & $-\frac{\sqrt{3}}{\sqrt{10}}$ & $-\frac{3}{\sqrt{10}}$ & $-\frac{3\sqrt{3}}{5\sqrt{2}}$ & $\frac{9\sqrt{6}}{5}$\\
         &$B^0_s\ra \eta_8\Kb$ & 0 & $-\frac{1}{\sqrt{10}}$ & $\frac{1}{\sqrt{30}}$ & $-\frac{1}{5\sqrt{2}}$ & $\frac{3\sqrt{2}}{5}$ & 0 & $-\frac{1}{\sqrt{10}}$ & $-\frac{\sqrt{3}}{\sqrt{10}}$ & $-\frac{3}{5\sqrt{2}}$ & $\frac{9\sqrt{2}}{5}$\\
         \hline\hline
         \multirow{12}{*}{$\Delta S = 1$}
         &$B^+\ra K^0\pi^+$ & 0 & $-\frac{\sqrt{3}}{\sqrt{5}}$ & $-\frac{1}{\sqrt{5}}$ & $\frac{3\sqrt{3}}{5}$ & $\frac{2\sqrt{3}}{5}$ & 0 & $-\frac{\sqrt{3}}{\sqrt{5}}$ & $\frac{3}{\sqrt{5}}$ & $\frac{9\sqrt{3}}{5}$ & $\frac{6\sqrt{3}}{5}$\\
         &$B^+\ra \pi^0K^+$ & 0 & $\frac{\sqrt{3}}{\sqrt{10}}$ &$\frac{1}{\sqrt{10}}$ & $-\frac{3\sqrt{3}}{5\sqrt{2}}$ & $\frac{4\sqrt{6}}{5}$ & 0 & $\frac{\sqrt{3}}{\sqrt{10}}$ & $-\frac{3}{\sqrt{10}}$ & $-\frac{9\sqrt{3}}{5\sqrt{2}}$ & $\frac{12\sqrt{6}}{5}$\\
         &$B^+\ra \eta_8 K^+$ & 0 & $-\frac{1}{\sqrt{10}}$ & $-\frac{1}{\sqrt{30}}$ & $\frac{3}{5\sqrt{2}}$ & $\frac{6\sqrt{2}}{5}$ & 0 & $-\frac{1}{\sqrt{10}}$ & $\frac{\sqrt{3}}{\sqrt{10}}$ & $\frac{9}{5\sqrt{2}}$ & $\frac{18\sqrt{2}}{5}$ \\
         \cline{2-12}
         &$B^0 \ra \pi^-K^+$ & 0 & $\frac{\sqrt{3}}{\sqrt{5}}$ & $-\frac{1}{\sqrt{5}}$ & $\frac{\sqrt{3}}{5}$ & $\frac{4\sqrt{3}}{5}$ & 0 & $\frac{\sqrt{3}}{\sqrt{5}}$ & $\frac{3}{\sqrt{5}}$ & $\frac{3\sqrt{3}}{5}$ & $\frac{12\sqrt{3}}{5}$\\
         &$B^0 \ra \pi^0K^0$ & 0 & $-\frac{\sqrt{3}}{\sqrt{10}}$ & $\frac{1}{\sqrt{10}}$ & $-\frac{\sqrt{3}}{5\sqrt{2}}$ & $\frac{3\sqrt{6}}{5}$ & 0 & $-\frac{\sqrt{3}}{\sqrt{10}}$ & $-\frac{3}{\sqrt{10}}$ & $-\frac{3\sqrt{3}}{5\sqrt{2}}$ & $\frac{9\sqrt{6}}{5}$\\
         &$B^0 \ra \eta_8 K^0$ & 0 & $-\frac{1}{\sqrt{10}}$ & $\frac{1}{\sqrt{30}}$ & $-\frac{1}{5\sqrt{2}}$ & $\frac{3\sqrt{2}}{5}$ & 0 & $-\frac{1}{\sqrt{10}}$ & $-\frac{\sqrt{3}}{\sqrt{10}}$ & $-\frac{3}{5\sqrt{2}}$ & $\frac{9\sqrt{2}}{5}$\\
         \cline{2-12}
         &$B^0_s \ra K^0\Kb$ & $-\frac{1}{2\sqrt{3}}$ & $-\frac{1}{\sqrt{15}}$ & $-\frac{1}{\sqrt{5}}$ & $-\frac{3\sqrt{3}}{5}$ & $\frac{\sqrt{3}}{10}$ & $-\frac{1}{2\sqrt{3}}$ & $-\frac{1}{\sqrt{15}}$ & $\frac{3}{\sqrt{5}}$ & $-\frac{9\sqrt{3}}{5}$ & $\frac{3\sqrt{3}}{10}$\\
         &$B^0_s \ra \pi^+\pi^-$ & $\frac{1}{2\sqrt{3}}$ & $-\frac{2}{\sqrt{15}}$ & 0 & $-\frac{2\sqrt{3}}{5}$ & $-\frac{\sqrt{3}}{10}$ & $\frac{1}{2\sqrt{3}}$ & $-\frac{2}{\sqrt{15}}$ & 0 & $-\frac{6\sqrt{3}}{5}$ & $-\frac{3\sqrt{3}}{10}$\\
         &$B^0_s \ra K^-K^+$ & $\frac{1}{2\sqrt{3}}$ & $\frac{1}{\sqrt{15}}$ & $-\frac{1}{\sqrt{5}}$ & $-\frac{\sqrt{3}}{5}$ & $\frac{7\sqrt{3}}{10}$ & $\frac{1}{2\sqrt{3}}$ & $\frac{1}{\sqrt{15}}$ & $\frac{3}{\sqrt{5}}$ & $-\frac{3\sqrt{3}}{5}$ & $\frac{21\sqrt{3}}{10}$\\
         &$B^0_s \ra \pi^0\pi^0$ & $-\frac{1}{2\sqrt{6}}$ & $\frac{\sqrt{2}}{\sqrt{15}}$ & 0 & $\frac{\sqrt{6}}{5}$ & $\frac{\sqrt{3}}{10\sqrt{2}}$ & $-\frac{1}{2\sqrt{6}}$ & $\frac{\sqrt{2}}{\sqrt{15}}$ & 0 & $\frac{3\sqrt{6}}{5}$ & $\frac{3\sqrt{3}}{10\sqrt{2}}$\\
         &$B^0_s \ra \pi^0 \eta_8$ & 0 & 0 & $\frac{2}{\sqrt{15}}$ & $\frac{4}{5}$ & $\frac{6}{5}$ & 0 & 0 & $-\frac{2\sqrt{3}}{\sqrt{5}}$ & $\frac{12}{5}$ & $\frac{18}{5}$\\
         &$B^0_s \ra \eta_8\eta_8$ & $-\frac{1}{2\sqrt{6}}$ & $-\frac{\sqrt{2}}{\sqrt{15}}$ & 0 & $-\frac{\sqrt{6}}{5}$ & $\frac{9\sqrt{3}}{10\sqrt{2}}$ & $-\frac{1}{2\sqrt{6}}$ & $-\frac{\sqrt{2}}{\sqrt{15}}$ & 0 & $-\frac{3\sqrt{6}}{5}$ & $\frac{27\sqrt{3}}{10\sqrt{2}}$\\
         \hline
    \end{tabular}
\end{center}
\caption{RME contributions to the $\Delta S = 0$ and $\Delta S=1$ $B \to PP$ amplitudes for the $(\boldsymbol{8\otimes 8})_S$ final state. \label{tab:8x8RME}}
\end{table}

\begin{table}[H]
\begin{center}
    \begin{tabular}{|l|l|c|c|c||c|c|c|}
         \hline
         \multicolumn{2}{|c|}{\multirow{3}{*}{Decays}} & \multicolumn{3}{c||}{$\lambda_u^{(q)}$} &\multicolumn{3}{c|}{$\lambda_t^{(q)}$}  \\
         \cline{3-8}
         \multicolumn{2}{|c|}{}&\multicolumn{3}{c||}{$c_1,c_2$}&$c_3$, $c_4$, $c_5$, $c_6$, $c_9$, $c_{10}$&\multicolumn{2}{c|}{$c_9$, $c_{10}$}\\
         \cline{3-8}
         \multicolumn{2}{|c|}{}& $C_8$ & $L_8$ & $M_8$ & $D_8$ & $L_8$ & $M_8$\\
         \hline\hline
         \multirow{4}{*}{$\Delta S = 0$}
         &$B^+ \ra \pi^+\eta_1$  & 1 & $\frac{1}{\sqrt{3}}$ & $-\frac{3}{\sqrt{5}}$ & $1$ & $-\sqrt{3}$ & $-\frac{9}{\sqrt{5}}$\\
         \cline{2-8}
         &$B^0 \ra \pi^0\eta_1$  & $\frac{1}{\sqrt{2}}$ & $\frac{1}{\sqrt{6}}$ & $\frac{\sqrt{5}}{\sqrt{2}}$ & $\frac{1}{\sqrt{2}}$ & $-\frac{\sqrt{3}}{\sqrt{2}}$ & $\frac{3\sqrt{5}}{\sqrt{2}}$\\
         &$B^0 \ra \eta_1\eta_8$  & $-\frac{1}{\sqrt{6}}$ & $\frac{1}{\sqrt{2}}$ & $\frac{\sqrt{3}}{\sqrt{10}}$ & $-\frac{1}{\sqrt{6}}$ & $-\frac{3}{\sqrt{2}}$ & $\frac{3\sqrt{3}}{\sqrt{10}}$\\
         \cline{2-8}
         &$B^0_s \ra \eta_1 \Kb$  & 1 & $-\frac{1}{\sqrt{3}}$ & $\frac{1}{\sqrt{5}}$ & $1$ & $\sqrt{3}$ & $\frac{3}{\sqrt{5}}$\\
         \hline\hline
         \multirow{4}{*}{$\Delta S=1$}
         &$B^+ \ra \eta_1 K^+$ & 1 & $\frac{1}{\sqrt{3}}$ & $-\frac{3}{\sqrt{5}}$ & $1$ & $-\sqrt{3}$ & $-\frac{9}{\sqrt{5}}$ \\
         \cline{2-8}
         &$B^0 \ra \eta_1 K^0$ & 1 & $-\frac{1}{\sqrt{3}}$ & $\frac{1}{\sqrt{5}}$ & $1$ & $\sqrt{3}$ & $\frac{3}{\sqrt{5}}$\\
         \cline{2-8}
         &$B^0_s \ra \pi^0 \eta_1$ & 0 & $\frac{\sqrt{2}}{\sqrt{3}}$ & $\frac{2\sqrt{2}}{\sqrt{5}}$ & 0 & $-\sqrt{6}$ & $\frac{6\sqrt{2}}{\sqrt{5}}$\\
         &$B^0_s \ra \eta_1\eta_8$ & $\frac{\sqrt{2}}{\sqrt{3}}$ & 0 & $\frac{\sqrt{6}}{\sqrt{5}}$ & $\frac{\sqrt{2}}{\sqrt{3}}$ & 0 & $\frac{3\sqrt{6}}{\sqrt{5}}$\\
         \hline
    \end{tabular}
\end{center}    
\caption{RME contributions to the $\Delta S = 0$ and $\Delta S=1$ $B \to PP$ amplitudes for the $\boldsymbol{8\otimes 1}$ final state.
\label{tab:8x1RME}}
\end{table}

\begin{table}[H]
\begin{center}    
\begin{tabular}{|l|l|c||c|} \hline
        \multicolumn{2}{|c|}{\multirow{2}{*}{Decays}} & $\lambda_u^{(q)}$ & $\lambda_t^{(q)}$\\
        \cline{3-4}
        \multicolumn{2}{|c|}{} &$C_1$ & $D_1$\\
        \hline\hline
        $\Delta S =0$ & $B^0\ra \eta_1\eta_1$ &$\frac{1}{\sqrt{3}}$ & $\frac{1}{\sqrt{3}}$ \\
        \hline\hline
        $\Delta S=1$ &$B^0_s\ra\eta_1\eta_1$ & $\frac{1}{\sqrt{3}}$ & $\frac{1}{\sqrt{3}}$\\
        \hline
\end{tabular}
\end{center}
\caption{RME contributions to the $\Delta S = 0$ and $\Delta S=1$ $B \to PP$ amplitudes for the $\boldsymbol{1\otimes 1}$ final state.    
\label{tab:1x1RME}}
\end{table}

\section{EWP diagrams}
\label{app:EWPs}

It is also possible to express the $B \to PP$ amplitudes in terms of topological quark diagrams, see the discussion in Sec.~\ref{Sec:diagrams}. The standard tree and penguin topologies can be found in Ref.~\cite{Gronau:1994rj}, for example. There are also six electroweak penguin topologies: two were introduced in Ref.~\cite{Gronau:1995hn}; the other four were later pointed out in Ref.~\cite{Gronau:1998fn}. In Fig.~\ref{fig:EWP} below, we display the six electroweak penguin diagrams.

\begin{figure}[H]
\begin{center}
\begin{subfigure}{0.5\textwidth}
\begin{center}
\includegraphics[width=0.55\textwidth]{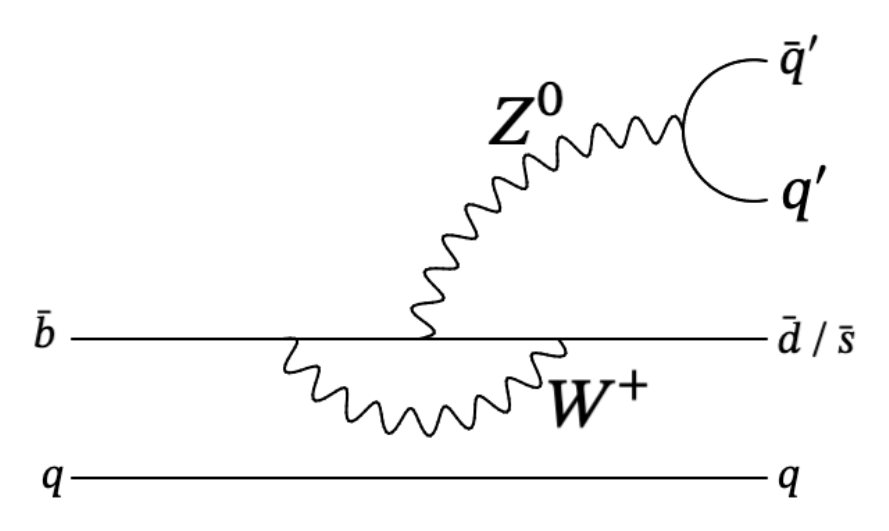}
\end{center}
   \caption{$P_{EW}^T$}
\end{subfigure}
\hspace{-1truecm}
\begin{subfigure}{0.5\textwidth}
\begin{center}
   \includegraphics[width=0.55\textwidth]{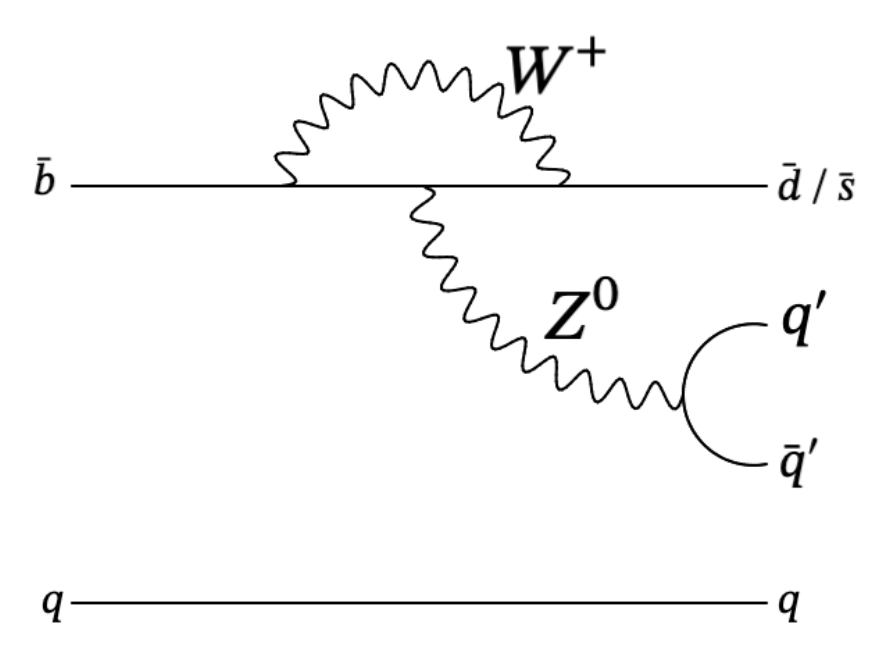}
\end{center}
   \caption{$P_{EW}^C$}
\end{subfigure}
\vskip 30pt

\begin{subfigure}{0.5\textwidth}
\begin{center}
   \includegraphics[width=0.55\textwidth]{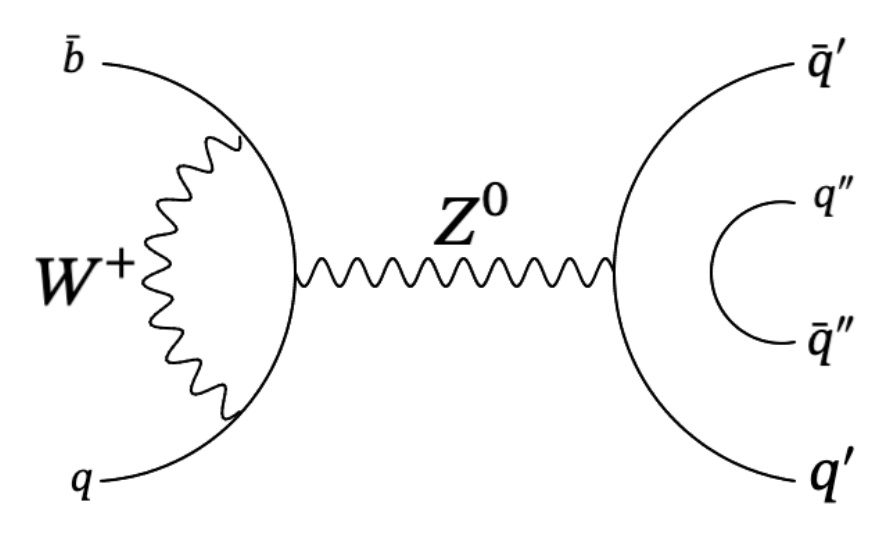}
\end{center}
   \caption{$P_{EW}^A$}
\end{subfigure}
\hspace{-1truecm}
\begin{subfigure}{0.5\textwidth}
\begin{center}
   \includegraphics[width=0.55\textwidth]{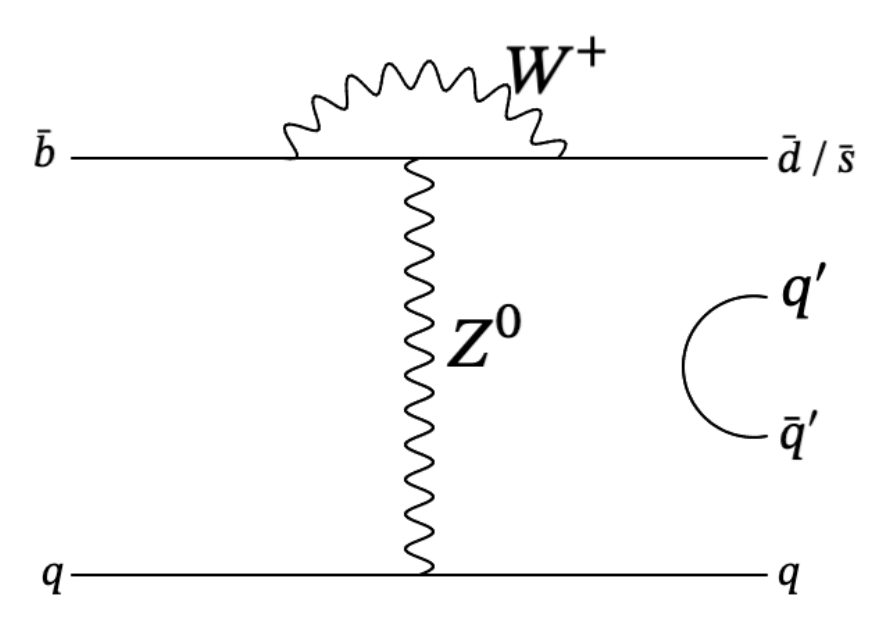}
\end{center}
   \caption{$P_{EW}^E$}
\end{subfigure}
\vskip 30pt

\begin{subfigure}{0.5\textwidth}
\begin{center}
   \includegraphics[width=0.55\textwidth]{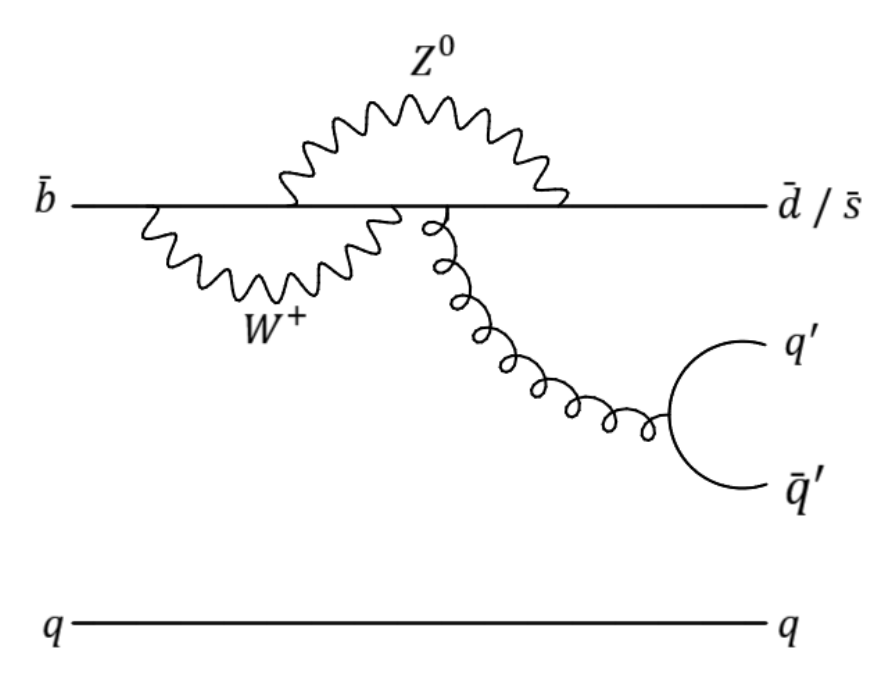}
\end{center}
   \caption{$P_{EW}^{P_u}$}
\end{subfigure}
\hspace{-1truecm}
\begin{subfigure}{0.5\textwidth}
\begin{center}
   \includegraphics[width=0.55\textwidth]{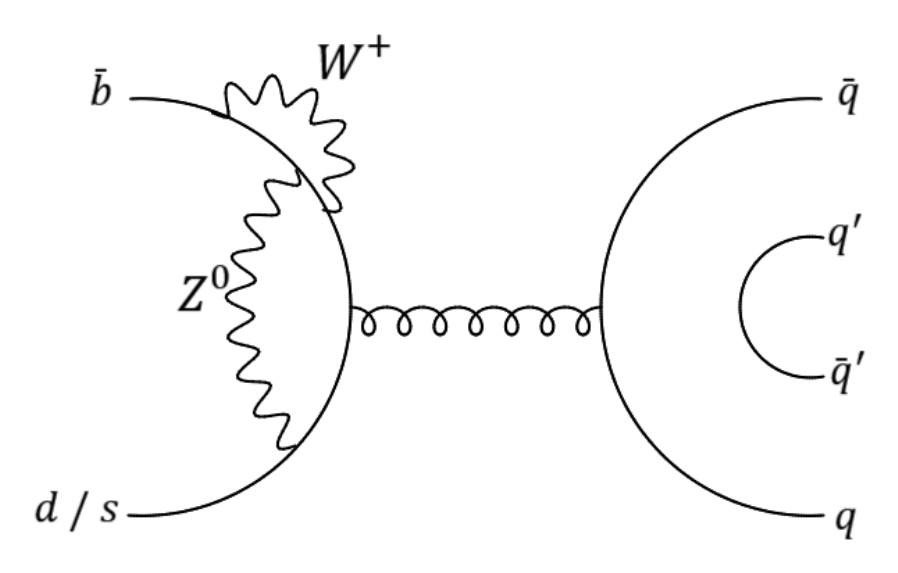}
\end{center}
   \caption{$P_{EW}^{PA_u}$}
\end{subfigure}
\vskip 30pt
\end{center}
\caption{Electroweak penguin diagrams contributing to the $B\ra PP$ decays
\label{fig:EWP}}
\end{figure}

\section{$B \to PP$ amplitudes: decomposition in terms of diagrams}
\label{app:diags}

The $B \to PP$ decay amplitudes can also be decomposed in terms of diagrams. For the three categories of decays, $(\boldsymbol{8\otimes 8})_S$, $\boldsymbol{8\otimes1}$, or $\boldsymbol{1\otimes1}$, this decomposition is shown below in Tables \ref{tab:8x8DIAG}, \ref{tab:8x1DIAG} and \ref{tab:1x1DIAG}, respectively. Note that, even though we use the same symbols for the diagrams in the $(\boldsymbol{8\otimes 8})_S$, $\boldsymbol{8\otimes 1}$ and $\boldsymbol{1\otimes 1}$ decay amplitudes ($T$, $C$, etc.), they are different in the three  decay categories.

\begin{table}[H]
\begin{center}
    \centering
    \begin{tabular}{|l|l|c|c|c|c|c|c||c|c|c|c|c|c|c|c|}
         \hline
         \multicolumn{2}{|c}{\multirow{2}{*}{Decays}} & \multicolumn{6}{|c||}{$\lambda_u^{(q)}$} & \multicolumn{8}{c|}{$\lambda_t^{(q)}$}\\
         \cline{3-16}
         \multicolumn{2}{|c|}{}& $T$ & $C$ & $P_{uc}$ & $A$ & $PA_{uc}$ & $E$ & $P_{tc}$ & $PA_{tc}$ & $P^T_{EW}$ & $P^C_{EW}$ & $P^A_{EW}$ & $P^E_{EW}$ & $P^{P_{u}}_{EW}$ & $P^{PA_{u}}_{EW}$ \\
         \hline\hline
         \multirow{12}{*}{$\Delta S = 0$}&$B^+\ra\Kb K^+$& 0 & 0 & 1 & 1 & 0 & 0 & 1 & 0 & 0 & $-\frac{1}{3}$ & 0 & $\frac{2}{3}$ & $-\frac{1}{3}$ & 0\\
         &$B^+\ra\pi^0\pi^+$ & $-\frac{1}{\sqrt{2}}$ & $-\frac{1}{\sqrt{2}}$ & 0 & 0 & 0 & 0 & 0 & 0 & $-\frac{1}{\sqrt{2}}$ & $-\frac{1}{\sqrt{2}}$ & 0 & 0 & 0 & 0\\
         &$B^+\ra\eta_8\pi^+$ & $-\frac{1}{\sqrt{6}}$ & $-\frac{1}{\sqrt{6}}$ & $-\frac{2}{\sqrt{6}}$ & $-\frac{2}{\sqrt{6}}$ & 0 & 0 & $-\frac{2}{\sqrt{6}}$ & 0 & $-\frac{1}{\sqrt{6}}$ & $-\frac{1}{3\sqrt{6}}$ & 0 & $-\frac{4}{3\sqrt{6}}$ & $\frac{2}{3\sqrt{6}}$ & 0\\
         \cline{2-16}
         &$B^0\ra K^0\Kb$ & 0 & 0 & 1 & 0 & 1 & 0 & 1 & 1 & 0 & $-\frac{1}{3}$ & $-\frac{2}{3}$ & $-\frac{1}{3}$ & $-\frac{1}{3}$ & $-\frac{1}{3}$\\
         &$B^0\ra\pi^+\pi^-$ & $-1$ & 0 & $-1$ & 0 & $-1$ & $-1$ & $-1$ & $-1$ & 0 & $-\frac{2}{3}$ & $-\frac{1}{3}$ & $\frac{1}{3}$ & $\frac{1}{3}$ & $\frac{1}{3}$\\
         &$B^0\ra K^+K^-$ & 0 & 0 & 0 & 0 & $-1$ & $-1$ & 0 & $-1$ & 0 & 0 & $-\frac{1}{3}$ & 0 & 0 & $\frac{1}{3}$\\
         &$B^0\ra\pi^0\pi^0$ & 0 & $-\frac{1}{\sqrt{2}}$ & $\frac{1}{\sqrt{2}}$ & 0 & $\frac{1}{\sqrt{2}}$ & $\frac{1}{\sqrt{2}}$ & $\frac{1}{\sqrt{2}}$ & $\frac{1}{\sqrt{2}}$ & $-\frac{1}{\sqrt{2}}$ & $-\frac{1}{3\sqrt{2}}$ & $\frac{1}{3\sqrt{2}}$ & $-\frac{1}{3\sqrt{2}}$ & $-\frac{1}{3\sqrt{2}}$ & $-\frac{1}{3\sqrt{2}}$\\
         &$B^0\ra\pi^0\eta_8$ & 0 & 0 & $-\frac{1}{\sqrt{3}}$ & 0 & 0 & $\frac{1}{\sqrt{3}}$ & $-\frac{1}{\sqrt{3}}$ & 0 & 0 & $\frac{1}{3\sqrt{3}}$ & $\frac{1}{\sqrt{3}}$ & $\frac{1}{3\sqrt{3}}$ & $\frac{1}{3\sqrt{3}}$ & 0 \\
         &$B^0\ra\eta_8\eta_8$ & 0 & $\frac{1}{3\sqrt{2}}$ & $\frac{1}{3\sqrt{2}}$ & 0 & $\frac{1}{\sqrt{2}}$ & $\frac{1}{3\sqrt{2}}$ & $\frac{1}{3\sqrt{2}}$ & $\frac{1}{\sqrt{2}}$ & $\frac{1}{3\sqrt{2}}$ & $-\frac{1}{9\sqrt{2}}$ & $-\frac{1}{3\sqrt{2}}$ & $-\frac{1}{9\sqrt{2}}$ & $-\frac{1}{9\sqrt{2}}$ & $-\frac{1}{3\sqrt{2}}$\\
         \cline{2-16}
         &$B^0_s\ra\pi^+K^-$ & $-1$ & 0 & $-1$ & 0 & 0 & 0 & $-1$ & 0 & 0 & $-\frac{2}{3}$ & 0 & $\frac{1}{3}$ & $\frac{1}{3}$ & 0\\
         &$B^0_s\ra\pi^0\Kb$ & 0  & $-\frac{1}{\sqrt{2}}$ & $\frac{1}{\sqrt{2}}$ & 0 & 0 & 0 & $\frac{1}{\sqrt{2}}$ & 0 & $-\frac{1}{\sqrt{2}}$ & $-\frac{1}{3\sqrt{2}}$ & 0 & $-\frac{1}{3\sqrt{2}}$ & $-\frac{1}{3\sqrt{2}}$ & 0\\
         &$B^0_s\ra\eta_8\Kb$ & 0 & $-\frac{1}{\sqrt{6}}$ & $\frac{1}{\sqrt{6}}$ & 0 & 0 & 0 & $\frac{1}{\sqrt{6}}$ & 0 & $-\frac{1}{\sqrt{6}}$ & $-\frac{1}{3\sqrt{6}}$ & 0 & $-\frac{1}{3\sqrt{6}}$ & $-\frac{1}{3\sqrt{6}}$ & 0\\
         \hline\hline
         \multirow{12}{*}{$\Delta S = 1$}
         &$B^+\ra \pi^+K^0$ & 0 & 0 & 1 & 1 & 0 & 0 & 1 & 0 & 0 & $-\frac{1}{3}$ & 0 & $\frac{2}{3}$ & $-\frac{1}{3}$ & 0\\
         &$B^+\ra\pi^0K^+$ & $-\frac{1}{\sqrt{2}}$ & $-\frac{1}{\sqrt{2}}$ & $-\frac{1}{\sqrt{2}}$ &  $-\frac{1}{\sqrt{2}}$ & 0 & 0 &$-\frac{1}{\sqrt{2}}$ & 0 & $-\frac{1}{\sqrt{2}}$ & $-\frac{\sqrt{2}}{3}$ & 0 & $-\frac{\sqrt{2}}{3}$ & $\frac{1}{3\sqrt{2}}$ & 0\\
         &$B^+\ra\eta_8K^+$ & $-\frac{1}{\sqrt{6}}$ & $-\frac{1}{\sqrt{6}}$ & $\frac{1}{\sqrt{6}}$ & $\frac{1}{\sqrt{6}}$ & 0 & 0 & $\frac{1}{\sqrt{6}}$ & 0 & $-\frac{1}{\sqrt{6}}$ & $-\frac{4}{3\sqrt{6}}$ & 0 & $\frac{2}{3\sqrt{6}}$ & $-\frac{1}{3\sqrt{6}}$ & 0\\
         \cline{2-16}
         &$B^0\ra\pi^-K^+$ & $-1$ & 0 & $-1$ & 0 & 0 & 0 & $-1$ & 0 & 0 & $-\frac{2}{3}$ & 0 & $\frac{1}{3}$ & $\frac{1}{3}$ & 0\\
         &$B^0\ra\pi^0K^0$ & 0 & $-\frac{1}{\sqrt{2}}$ & $\frac{1}{\sqrt{2}}$ &  0 & 0 & 0 & $\frac{1}{\sqrt{2}}$ & 0 & $-\frac{1}{\sqrt{2}}$ & $-\frac{1}{3\sqrt{2}}$ & 0 & $-\frac{1}{3\sqrt{2}}$ & $-\frac{1}{3\sqrt{2}}$ & 0\\
         &$B^0\ra\eta_8K^0$ & 0 & $-\frac{1}{\sqrt{6}}$ & $\frac{1}{\sqrt{6}}$ &  0 & 0 & 0 & $\frac{1}{\sqrt{6}}$ & 0 & $-\frac{1}{\sqrt{6}}$ & $-\frac{1}{3\sqrt{6}}$ & 0 & $-\frac{1}{3\sqrt{6}}$ & $-\frac{1}{3\sqrt{6}}$ & 0\\
         \cline{2-16}
         &$B^0_s\ra K^0\Kb$ & 0 & 0 & 1 & 0 & 1 & 1 & 1 & 1 & 0 & $-\frac{1}{3}$ & $-\frac{2}{3}$ & $-\frac{1}{3}$ & $-\frac{1}{3}$ & $-\frac{1}{3}$\\
         &$B^0_s\ra\pi^+\pi^-$ & 0 & 0 & 0 & 0 & $-1$ & $-1$ & 0 & $-1$ & 0 & 0 & $-\frac{1}{3}$ & 0 & 0 & $\frac{1}{3}$\\
         &$B^0_s\ra K^+K^-$ & $-1$ & 0 & $-1$ & 0 & $-1$ & $-1$ & $-1$ & $-1$ & 0 & $-\frac{2}{3}$ & $-\frac{1}{3}$ & $\frac{1}{3}$ & $\frac{1}{3}$ & $\frac{1}{3}$\\
         &$B^0_s\ra\pi^0\pi^0$ & 0 & 0 & 0 & 0 & $\frac{1}{\sqrt{2}}$ & $\frac{1}{\sqrt{2}}$ & 0 & $\frac{1}{\sqrt{2}}$ & 0 & 0 & $\frac{1}{3\sqrt{2}}$ & 0 & 0 & $-\frac{1}{3\sqrt{2}}$\\
         &$B^0_s\ra\pi^0\eta_8$ & 0 & $-\frac{1}{\sqrt{3}}$ & 0 & 0 & 0 & $\frac{1}{\sqrt{3}}$ & 0 & 0 & $-\frac{1}{\sqrt{3}}$ & 0 & $\frac{1}{\sqrt{3}}$ & 0 & 0 & 0\\
         &$B^0_s\ra\eta_8\eta_8$ & 0 & $-\frac{\sqrt{2}}{3}$ & $\frac{2\sqrt{2}}{3}$ & 0 & $\frac{1}{\sqrt{2}}$ & $\frac{1}{3\sqrt{2}}$ & $\frac{2\sqrt{2}}{3}$ & $\frac{1}{\sqrt{2}}$ & $-\frac{\sqrt{2}}{3}$ & $-\frac{2\sqrt{2}}{9}$ & $-\frac{1}{3\sqrt{2}}$ & $-\frac{2\sqrt{2}}{9}$ & $-\frac{2\sqrt{2}}{9}$ & $-\frac{1}{3\sqrt{2}}$\\
         \hline
    \end{tabular}
\end{center}
\caption{Diagrammatic contributions to the $\Delta S = 0$ and $\Delta S=1$ $B \to PP$ amplitudes for the $(\boldsymbol{8\otimes 8})_S$ final state. \label{tab:8x8DIAG}}
\end{table}

\begin{table}[H]
\begin{center}
\begin{tabular}{|l|l|c|c|c|c|c|c||c|c|c|c|c|c|c|c|}
         \hline
         \multicolumn{2}{|c|}{\multirow{2}{*}{Decays}} & \multicolumn{6}{c||}{$\lambda_u^{(q)}$} & \multicolumn{8}{c|}{$\lambda_t^{(q)}$}\\
         \cline{3-16}\multicolumn{2}{|c|}{}& $T$ & $C$ & $P_{uc}$ & $A$ & $PA_{uc}$ & $E$ & $P_{tc}$ & $PA_{tc}$ & $P^T_{EW}$ & $P^C_{EW}$ & $P^A_{EW}$ & $P^E_{EW}$ & $P^{P_{u}}_{EW}$ & $P^{PA_{u}}_{EW}$\\
         \hline\hline
         \multirow{4}{*}{$\Delta S=0$}
         &$B^+\ra\pi^+\eta_1$ & $\frac{1}{\sqrt{3}}$ & $\frac{1}{\sqrt{3}}$ & $\frac{2}{\sqrt{3}}$ & $\frac{2}{\sqrt{3}}$ & 0 & 0 &$\frac{2}{\sqrt{3}}$ & 0 & 0 & $\frac{1}{3\sqrt{3}}$ & 0 & $\frac{4}{3\sqrt{3}}$ & $-\frac{2}{3\sqrt{3}}$ & 0 \\
         \cline{2-16}
         &$B^0\ra\pi^0\eta_1$ & 0 & 0 & $\frac{2}{\sqrt{6}}$ & 0 & 0 &$-\frac{2}{\sqrt{6}}$ & $\frac{2}{\sqrt{6}}$ & 0 & $-\frac{1}{\sqrt{6}}$ & $-\frac{2}{3\sqrt{6}}$ & $-\frac{2}{\sqrt{6}}$ & $-\frac{2}{3\sqrt{6}}$ & $-\frac{2}{3\sqrt{6}}$ & 0 \\
         &$B^0\ra\eta_1\eta_8$ & 0 & $-\frac{\sqrt{2}}{3}$ & $-\frac{\sqrt{2}}{3}$ & 0 & 0 & $-\frac{\sqrt{2}}{3}$ & $-\frac{\sqrt{2}}{3}$ & 0 & $-\frac{1}{3\sqrt{2}}$ & $\frac{\sqrt{2}}{9}$ & $-\frac{\sqrt{2}}{3}$ & $\frac{\sqrt{2}}{9}$ & $\frac{\sqrt{2}}{9}$ & 0 \\
         \cline{2-16}
         &$B^0_s\ra\eta_1\Kb$ & 0 & $\frac{1}{\sqrt{3}}$ & $\frac{2}{\sqrt{3}}$ & 0 & 0 & 0 & $\frac{2}{\sqrt{3}}$ & 0 & 0 & $-\frac{2}{3\sqrt{3}}$ & 0 & $-\frac{2}{3\sqrt{3}}$ & $-\frac{2}{3\sqrt{3}}$ & 0 \\
         \hline\hline
         \multirow{4}{*}{$\Delta S=1$}
         & $B^+\ra \eta_1 K^+$ & $\frac{1}{\sqrt{3}}$ & $\frac{1}{\sqrt{3}}$ & $\frac{2}{\sqrt{3}}$ & $\frac{2}{\sqrt{3}}$ & 0 & 0 & $\frac{2}{\sqrt{3}}$ & 0 & 0 & $\frac{1}{3\sqrt{3}}$ & 0 & $\frac{4}{3\sqrt{3}}$ & $-\frac{2}{3\sqrt{3}}$ & 0 \\
         \cline{2-16}
         & $B^0\ra\eta_1K^0$ & 0 & $\frac{1}{\sqrt{3}}$ & $\frac{2}{\sqrt{3}}$ & 0 & 0 & 0 & $\frac{2}{\sqrt{3}}$ & 0 & 0 & $-\frac{2}{3\sqrt{3}}$ & 0 & $-\frac{2}{3\sqrt{3}}$ & $-\frac{2}{3\sqrt{3}}$ & 0 \\
         \cline{2-16}
         & $B^0_s\ra\pi^0\eta_1$ & 0 & $-\frac{1}{\sqrt{6}}$ & 0 & 0 & 0 & $-\frac{2}{\sqrt{6}}$ & 0 & 0 & $-\frac{1}{\sqrt{6}}$ & 0 & $-\frac{2}{\sqrt{6}}$ & 0 & 0 & 0\\
         &$B^0_s\ra\eta_8\eta_1$ & 0 & $\frac{1}{3\sqrt{2}}$ & $\frac{2\sqrt{2}}{3}$ & 0 & 0 & $-\frac{2}{3\sqrt{2}}$ & $\frac{2\sqrt{2}}{3}$ & 0 & $-\frac{1}{3\sqrt{2}}$ & $-\frac{2\sqrt{2}}{9}$ & $-\frac{\sqrt{2}}{3}$ & $-\frac{2\sqrt{2}}{9}$ & $-\frac{2\sqrt{2}}{9}$ & 0 \\
         \hline
\end{tabular}
\end{center}
\caption{Diagrammatic contributions to the $\Delta S = 0$ and $\Delta S=1$ $B \to PP$ amplitudes for the $\boldsymbol{8\otimes 1}$ final state. \label{tab:8x1DIAG}}
\end{table}

\begin{table}[H]
\begin{center}
    \begin{tabular}{|l|l|c|c|c|c|c|c||c|c|c|c|c|c|c|c|}
        \hline
        \multicolumn{2}{|c|}{\multirow{2}{*}{Decays}} & \multicolumn{6}{c||}{$\lambda_u^{(q)}$} & \multicolumn{8}{c|}{$\lambda_t^{(q)}$}\\
        \cline{3-16}
        \multicolumn{2}{|c|}{}& $T$ & $C$ & $P_{uc}$ & $A$ & $PA_{uc}$ & $E$ & $P_{tc}$ & $PA_{tc}$ & $P^T_{EW}$ & $P^C_{EW}$ & $P^A_{EW}$ & $P^E_{EW}$ & $P^{P_{u}}_{EW}$ & $P^{PA_{u}}_{EW}$ \\
        \hline\hline
        $\Delta S = 0$ & $B^0\ra\eta_1\eta_1$ & 0 & $\frac{\sqrt{2}}{3}$ & $\frac{\sqrt{2}}{3}$ & 0 & $2\sqrt{2}$ & $\frac{\sqrt{2}}{3}$ & $\frac{\sqrt{2}}{3}$ & $2\sqrt{2}$ & 0 & $-\frac{\sqrt{2}}{9}$ & 0 & $-\frac{\sqrt{2}}{9}$ & $-\frac{\sqrt{2}}{9}$ & $-\frac{2\sqrt{2}}{3}$\\
        \hline\hline
        $\Delta S = 1$ & $B^0_s\ra\eta_1\eta_1$ & 0 & $\frac{\sqrt{2}}{3}$ & $\frac{\sqrt{2}}{3}$ & 0 & $2\sqrt{2}$ & $\frac{\sqrt{2}}{3}$ & $\frac{\sqrt{2}}{3}$ & $2\sqrt{2}$ & 0 & $-\frac{\sqrt{2}}{9}$ & 0 & $-\frac{\sqrt{2}}{9}$ & $-\frac{\sqrt{2}}{9}$ & $-\frac{2\sqrt{2}}{3}$\\
        \hline
    \end{tabular}
\end{center}
\caption{Diagrammatic contributions to the $\Delta S = 0$ and $\Delta S=1$ $B \to PP$ amplitudes for the $\boldsymbol{1\otimes 1}$ final state. \label{tab:1x1DIAG}}
\end{table}

\section{Measured $B \to PP$ observables}
\label{app:data}

Here we present the data used in our analysis. In Table \ref{tab:exp_data}, we list the values of CP-averaged branching ratios (${\cal B}_{CP}$), direct CP asymmetries ($A_{CP}$), and indirec CP asymmetries ($S_{CP}$) used in our fits. 

\begin{table}[H]
\begin{center}
    \begin{tabular}{|l |l|c|c|c|}
        \hline
        \multicolumn{2}{|l|}{Decays} & ${\cal B}_{CP}$ ($\times 10^{-6}$) & $A_{CP}$ & $S_{CP}$ \\
        \hline\hline
        \multirow{17}{*}{$\Delta S = 0$}
        &$B^+\ra\Kb K^+$ & $1.31\pm0.14^*$& $0.04\pm0.14$ & \\
        &$B^+\ra\pi^0\pi^+$ & $5.31\pm0.26$ & $-0.01\pm0.04$ & \\
        &$B^+\ra\eta \pi^+$ & $4.02\pm0.27$ & $-0.14\pm0.07$ & \\
        &$B^+\ra\eta' \pi^+$ & $2.7\pm0.9$ & $0.06\pm0.16$ & \\
        \cline{2-5}
        &$B^0\ra K^0\Kb$ & $1.21\pm0.16$ & $0.06\pm0.26^*$ & $-1.08\pm0.49^*$ \\
        &$B^0\ra\pi^+\pi^-$ & $5.43\pm0.26$ & $0.314\pm0.030$ & $-0.670\pm0.030$ \\
        &$B^0\ra K^+K^-$ & $0.082\pm0.015^*$ & & \\
        &$B^0\ra\pi^0\pi^0$ & $1.55\pm0.17$ & $0.30\pm0.20$ & \\
        &$B^0\ra\pi^0\eta$ & $0.41\pm0.17$ & & \\
        &$B^0\ra\pi^0\eta'$ & $1.2\pm0.6$ & & \\
        &$B^0\ra\eta\eta$ & $0.5\pm0.3\pm0.1^{\dagger}$ & & \\
        &$B^0\ra\eta\eta'$ & & & \\
        &$B^0\ra\eta'\eta'$ & $0.6^{+0.5}_{-0.4}\pm0.4^{\dagger}$ & & \\
        \cline{2-5}
        &$B^0_s\ra\pi^+K^-$ & $5.9\pm0.7$ & $0.224\pm0.012$ & \\
        &$B^0_s\ra\pi^0\Kb$ & & & \\
        &$B^0_s\ra\eta\Kb$ & & & \\
        &$B^0_s\ra\eta'\Kb$ & & & \\
        \hline\hline
        \multirow{17}{*}{$\Delta S = 1$}
        &$B^+\ra\pi^+K^0$ & $23.9\pm0.6$ & $-0.003\pm0.015$ & \\
        &$B^+\ra\pi^0K^+$ & $13.20\pm0.40$ & $0.027\pm0.012$ & \\
        &$B^+\ra\eta K^+$ & $2.4\pm0.4$ & $-0.37\pm0.08$ & \\
        &$B^+\ra\eta'K^+$ & $70.4\pm2.5$ & $0.004\pm0.011$ & \\
        \cline{2-5}
        &$B^0\ra\pi^-K^+$ & $20.0\pm0.4$ & $-0.0831\pm0.0031$ & \\
        &$B^0\ra\pi^0K^0$ & $10.1\pm0.4$ & $0.00\pm0.08$ & $0.64\pm0.13$ \\
        &$B^0\ra\eta K^0$ & $1.23^{+0.27}_{-0.24}$ & & \\
        &$B^0\ra\eta'K^0$ & $66\pm4$ & $0.06\pm0.04$ & $0.63\pm0.06$ \\
        \cline{2-5}
        &$B^0_s\ra K^0\Kb$ & $17.6\pm3.1$ & & \\
        &$B^0_s\ra\pi^+\pi^-$ & $0.72\pm0.10$ & & \\
        &$B^0_s\ra K^+K^-$ & $27.2\pm2.3$ & $-0.162\pm0.035$ & $0.14\pm0.03^*$ \\
        &$B^0_s\ra\pi^0\pi^0$ & $2.8\pm2.8^{\dagger\dagger}$ & & \\
        &$B^0_s\ra\pi^0\eta$ & & & \\
        &$B^0_s\ra\pi^0\eta'$ & & & \\
        &$B^0_s\ra\eta\eta$ & $100\pm105\pm23^{\dagger\dagger\dagger}$ & & \\
        &$B^0_s\ra\eta\eta'$ & $25\pm22\pm6^{\dagger\dagger\dagger\dagger}$ & & \\
        &$B^0_s\ra\eta'\eta'$ & $33\pm7\pm1$ & & \\
        \hline
    \end{tabular}
\end{center}
\caption{Measured observables in $B\to PP$ decays. Data labeled $^*,~^\dagger,~ ^{\dagger\dagger},~^{\dagger\dagger\dagger}$, and $^{\dagger\dagger\dagger\dagger}$ are taken from Refs.~\cite{HeavyFlavorAveragingGroupHFLAV:2024ctg} (\textit{HFLAV}), \cite{BaBar:2009cun}, \cite{Belle:2023aau}, \cite{Belle:2021tsq} and \cite{Belle:2021bxn}, respectively. Remaining data are taken from the Particle Data Group \cite{ParticleDataGroup:2024cfk} \label{tab:exp_data}}
\end{table}

\section{Fit results}
\label{app:res}

Here we present a complete list of our fit results for several fits performed in this paper. In Table \ref{tab:fitresults8x8}, the results are for three different fits involving final states with only pions and kaons (i.e., there is no $\eta$ or $\eta'$ in the final state). Table \ref{tab:fitresults} includes the results for six additional fits, three of which involve final states with at most one $\eta^{(\prime)}$, while the remaining three include all the $B\to PP$ data.

The results quoted in Tables \ref{tab:fitresults8x8} and \ref{tab:fitresults}, which include the central values and 1$\sigma$ Hessian error ranges, were obtained using the \textit{Minuit} package \cite{James:1975dr}. We have rounded the results to quote only one or two significant figures in the error. The $\chi^2_{\rm min}$ quoted in each column represents the minimum value of $\chi^2$ obtained through fitting and has been quoted up to three decimal places.

However, there is one extremely important point that the reader must be aware of: most of these fits have several parameters that are very highly correlated. As such, the value of the $\chi^2$ function itself is very sensitive to changes in the values of some of the fit parameters. Since we have rounded the results while quoting them, in many of these cases, simply using the central values as arguments in the $\chi^2$ function may not yield the $\chi^2_{\rm min}$ value quoted in that column. In these cases, the $\chi^2_{\rm min}$ will correspond to parameter values where additional digits after the decimal point are retained. We invite the interested reader to use the computer code accompanying this article \cite{Perollaz:2025code} to obtain the precise values of the parameters that reproduce each $\chi^2_{\rm min}$ value.

\begin{table}[H]
\begin{center}
\begin{tabular}{l|c|c|c} \hline\hline
\multicolumn{1}{c|}{Fit}             & \multicolumn{3}{c}{Type of fit} \\ \cline{2-4}
\multicolumn{1}{c|}{Parameters}      & $\Delta S = 0$ & $\Delta S = 1$ & Full fit         \\ \hline \hline
Re$({\tilde T})$ & $4.2\pm0.5$    & $55\pm6$       & $5.5\pm0.5$      \\
Re$({\tilde C})$ & $2.9\pm1.1$    & $-56\pm6$      & $2.47\pm0.20$    \\
Im$({\tilde C})$ & $-5.6\pm1.0$   & $-4.1\pm1.7$   & $-4.3\pm0.8$     \\
Re$({\tilde P_{uc}})$      
                 & $2.4\pm0.7$       & $-60\pm7$      & $1.4\pm0.8$      \\
Im$({\tilde P_{uc}})$      
                 & $-1\pm6$   & $-3.8\pm1.8$   & $-0.40\pm0.06$   \\
Re$({\tilde A})$ & $-0.1\pm0.8$        & $5\pm27$       & $-3.2\pm1.0$     \\
Im$({\tilde A})$ & $-0.2\pm1.1$        & $-30\pm40$     & $1.17\pm0.08$    \\
Re$(\widetilde {PA}_{uc})$ 
                 & $-0.6\pm0.5$       & $1\pm5$        & $-0.69\pm0.29$   \\
Im$(\widetilde {PA}_{uc})$ 
                 & $0\pm6$      & $-3.4\pm1.9$   & $-0.04\pm0.017$  \\
Re$({\tilde P_{tc}})$      
                 & $0.7\pm2.1$    & $-0.66\pm0.12$ & $1.09\pm0.11$    \\
Im$({\tilde P_{tc}})$
                 & $-0.42\pm0.31$   & $-0.3\pm0.4$   & $-0.45\pm0.05$   \\
Re$(\widetilde {PA}_{tc})$ 
                 & $0.1\pm2.1$    & $-0.18\pm0.12$ & $-0.079\pm0.020$ \\
Im$(\widetilde {PA}_{tc})$ 
                 & $-0.20\pm0.29$    & $-0.01\pm0.18$ & $-0.20\pm0.05$   \\ \hline \hline
$\chi^2_{\rm min}$  
                 & $1.057$        & $1.547$        & $43.19$          \\ \hline \hline
\end{tabular}
\end{center}
\caption{Fit results for $B\to PP$ decays, excluding final states containing an $\eta$ or $\eta'$, using $(\mathbf{8}\times\mathbf{8})_S$ diagrammatic amplitudes. Real and Imaginary parts of fit amplitudes are given in units of keV.
\label{tab:fitresults8x8}}
\end{table}
\clearpage

\begin{table}[H]
\begin{center}
\begin{tabular}{c|l|c|c|c|c|c|c} \hline \hline
\multicolumn{2}{c|}{Fit} & \multicolumn{3}{c|}{With one $\eta$ or $\eta^\prime$} & \multicolumn{3}{c}{All decays} \\ \cline{3-8}
\multicolumn{2}{c|}{Parameters} & $\Delta S = 0$ & $\Delta S = 1$ & Full fit & $\Delta S = 0$ & $\Delta S = 1$ & Full fit \\ \hline \hline
\multirow{13}{*}{\rotatebox{90}{$(\boldsymbol{8\otimes 8})_S$}} 
        & Re$({\tilde T})$ & $4.2\pm0.4$       & $55\pm6$      & $5.6\pm 0.5$      & $4.25\pm0.34$     & $53\pm6$          & $5.7\pm0.5$\\
        & Re$({\tilde C})$ & $2.9\pm0.9$       & $-56\pm6$     & $2.4\pm0.6$       & $2.9\pm0.7$       & $-54\pm7$         & $2.5\pm0.6$\\
        & Im$({\tilde C})$ & $-5.6\pm0.8$      & $-4.2\pm1.7$  & $-4.3\pm0.8$      & $-5.6\pm0.7$      & $-4.1\pm2.0$      & $-4.1\pm0.8$\\
        & Re$({\tilde P_{uc}})$ & $2.4\pm0.5$       & $-60\pm7$     & $1.3\pm0.5$       & $2.4\pm0.4$       & $-58\pm8$         & $1.3\pm0.5$\\
        & Im$({\tilde P_{uc}})$ & $-1.0\pm0.9$      & $-3.9\pm1.9$  & $-0.3\pm0.7$      & $-1.0\pm0.5$      & $-3.8\pm2.1$      & $-0.1\pm0.7$\\
        & Re$({\tilde A})$ & $-0.1\pm0.6$      & $6\pm30$      & $-2.9\pm0.6$      & $-0.1\pm0.5$      & $1\pm28$          & $-2.8\pm0.6$\\
        & Im$({\tilde A})$ & $-0.2\pm0.6$      & $-30\pm40$    & $0.9\pm0.8$       & $-0.2\pm0.5$      & $-30\pm50$        & $0.7\pm0.8$\\
        & Re$(\widetilde {PA}_{uc})$ & $-0.62\pm0.28$    & $1\pm5$       & $-0.68\pm0.10$    & $-0.62\pm0.22$    & $0\pm6$           & $-0.68\pm0.10$\\
        & Im$(\widetilde {PA}_{uc})$ & $-0.2\pm0.9$      & $-3.4\pm2.0$  & $-0.10\pm0.24$    & $-0.2\pm0.5$      & $-3.0\pm2.1$      & $-0.11\pm0.24$\\
        & Re$({\tilde P_{tc}})$ & $0.71\pm0.33$     & $-0.66\pm0.14$& $1.10\pm0.04$     & $0.71\pm0.18$     & $-0.68\pm0.12$    & $1.11\pm0.04$\\
        & Im$({\tilde P_{tc}})$ & $-0.42\pm0.09$    & $-0.3\pm0.4$  & $-0.44\pm0.10$    & $-0.42\pm0.06$    & $-0.3\pm0.5$      & $-0.41\pm0.10$\\
        & Re$(\widetilde {PA}_{tc})$ & $0.11\pm0.35$     & $-0.18\pm0.13$& $-0.08\pm0.05$    & $0.11\pm0.20$     & $-0.21\pm0.13$    & $-0.07\pm0.05$\\
        & Im$(\widetilde {PA}_{tc})$ & $-0.20\pm0.09$    & $0.00\pm0.18$ & $-0.203\pm0.023$  & $-0.20\pm0.07$    & $0.13\pm0.24$     & $-0.204\pm0.022$\\
        \hline \hline
        \multirow{8}{*}{\rotatebox{90}{$\boldsymbol{8\otimes1}$}}
        & Re$({\tilde T})$ & $7\pm9$        & $-176\pm12$   & $6\pm5$       & $-3\pm4$      & $-100\pm150$  & $8.5\pm3.3$\\
        & Im$({\tilde T})$ & $0\pm13$       & $-20\pm150$   & $4\pm5$       & $-1\pm7$      & $-160\pm90$   & $-2.2\pm3.3$\\
        & Re$({\tilde C})$ & $-1\pm9$       & $230\pm50$    & $5\pm5$       & $4\pm4$       & $160\pm60$    & $2.6\pm3.3$\\
        & Im$({\tilde C})$ & $11\pm13$      & $0\pm70$      & $-11\pm5$     & $0\pm7$       & $-10\pm40$    & $-2.4\pm2.9$\\
        & Re$({\tilde P_{uc}})$ & $-2.0\pm1.8$   & $-122\pm24$   & $-2.3\pm1.3$  & $0.6\pm1.6$   & $-89\pm30$    & $-2.5\pm1.3$\\
        & Im$({\tilde P_{uc}})$ & $-3.1\pm1.6$   & $0\pm40$      & $0.6\pm2.1$   & $3.4\pm1.1$   & $7\pm22$      & $-0.3\pm2.0$\\
        & Re$({\tilde P_{tc}})$ & $0.19\pm0.31$  & $-0.87\pm0.19$& $1.31\pm0.15$ & $0.18\pm0.25$ & $-1.0\pm0.6$  & $1.38\pm0.14$\\
        & Im$({\tilde P_{tc}})$ & $0.66\pm0.32$  & $0.0\pm0.4$   & $-1.26\pm0.15$& $0.69\pm0.28$ & $-0.9\pm0.7$  & $-1.19\pm0.16$\\
        \hline\hline
        \multirow{4}{*}{\rotatebox{90}{$\boldsymbol{1\otimes1}$}}
        & Re$({\tilde C})$ &&&& $-1\pm27$   & $0\pm700$     & $-1\pm9$\\
        & Im$({\tilde C})$ &&&& $0\pm210$   & $-100\pm800$  & $4\pm4$\\
        & Re$({\tilde P_{tc}})$ &&&& $-2\pm13$   & $-1\pm7$      & $0.7\pm2.3$\\
        & Im$({\tilde P_{tc}})$ &&&& $0\pm40$    & $1\pm18$      & $-0.3\pm3.1$\\
        \hline\hline
        \multicolumn{2}{c|}{$\chi^2_{\rm min}$}  & $1.057$ & $1.547$ & $55.62$     & $1.057$       & $2.437$   & $60.77$ \\
        \hline
\end{tabular}
\end{center}
\caption{Fit results for $B\to PP$ decays, including final states with at least one $\eta$ or $\eta'$. Real and Imaginary parts of fit amplitudes are given in units of keV.
\label{tab:fitresults}}
\end{table}

\end{document}